\documentclass[
  aps,prl,reprint,superscriptaddress,
  amsmath,amssymb,floatfix,nofootinbib
]{revtex4-2}
\usepackage{graphicx}
\usepackage{bm}
\usepackage[colorlinks=true,linkcolor=blue,citecolor=blue,urlcolor=blue]{hyperref}
\usepackage{fontawesome5}

\newcommand{\meV}{\,\mathrm{meV}}
\newcommand{\gagg}{g_{a\gamma\gamma}}
\newcommand{\Qm}{Q_{\rm m}}
\newcommand{\Qr}{Q_{\rm r}}
\newcommand{\dd}{\mathrm d}

\begin{document}

\title{Layered Wire-Grid Haloscope for Millielectronvolt Axions and Dark Photons}

\author{Andrea Caputo}
\affiliation{Department of Theoretical Physics, CERN, Esplanade des Particules 1, P.O. Box 1211, Geneva 23, Switzerland}
\affiliation{School of Natural Sciences, Institute for Advanced Study, Princeton, NJ 08540, USA}
\affiliation{Dipartimento di Fisica, ``Sapienza'' Universit\`a di Roma \& Sezione INFN Roma1, Piazzale Aldo Moro 5, 00185, Roma, Italy}

\date{\today}
\begin{abstract}
We propose MADGRID (\textbf{M}illielectronvolt \textbf{A}xions and \textbf{D}ark photons with layered \textbf{GRID}s), a few wire grids in front of a mirror, tuned by their spacing, forming an epsilon-near-zero slab that radiates through its face into external optics. Unlike wire-plasma haloscopes, the signal need not cross the lossy medium to reach an antenna, and metal grids replace the disks of dielectric haloscopes. A 1\,m$^2$ array in 10\,T could probe axion-photon couplings below the CAST bound up to 1.4\,meV with kinetic-inductance detectors, and KSVZ couplings over 0.8--3.9\,meV with future photon counters. Without a magnet, it could probe dark photon mixings down to $3.4\times10^{-16}$.~\href{https://github.com/andrea0292/Wiregrid_Haloscope}{\faGithub}
\end{abstract}

\maketitle

\begin{figure}[t]
\includegraphics[width=0.9\columnwidth]{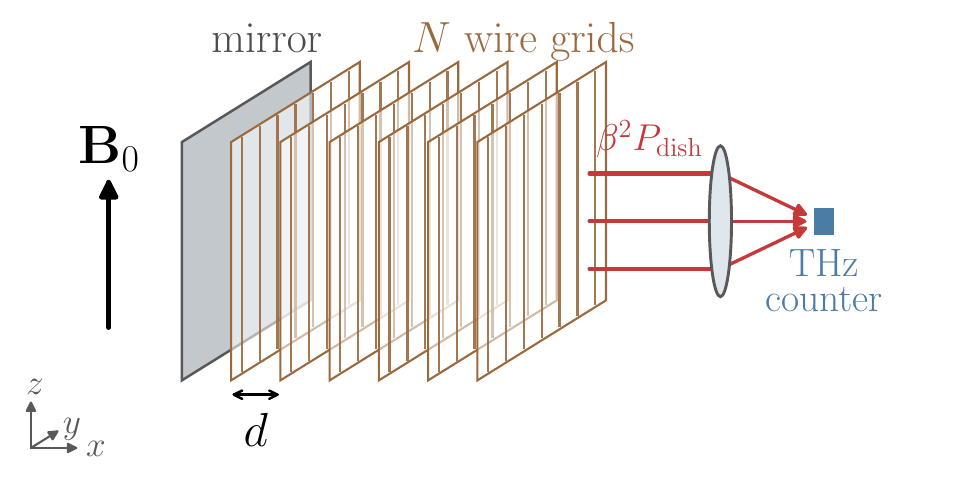}
\caption{A mirror and $N$ grids of wires parallel to $\mathbf B_0$, at spacing $d$, form an epsilon-near-zero (ENZ) slab. On resonance, it radiates through its face the power $|\beta|^2P_{\rm dish}$, with $P_{\rm dish}$ the emission of a bare mirror of the same area and $|\beta|^2$ the power boost. External optics focus it onto a THz detector.}
\label{fig:concept}
\end{figure}

\textit{Introduction.}---The QCD axion is a very well motivated new particle, introduced to explain why the strong interactions conserve CP~\cite{Peccei:1977hh,Weinberg:1977ma,Wilczek:1977pj}, and subsequently recognized as one of the most appealing dark matter (DM) candidates~\cite{Preskill:1982cy,Abbott:1982af,Dine:1982ah, OHare:2024nmr, Caputo:2024oqc, Safdi:2022xkm}. It couples to photons through the interaction $\mathcal L\supset\gagg a\,\mathbf E\cdot\mathbf B$, with a coupling proportional to its mass, $|\gagg|\simeq2.04\times10^{-13}|C_{a\gamma}|(m_a/{\rm meV})\,{\rm GeV}^{-1}$, where $|C_{a\gamma}|=1.92$ in the KSVZ model~\cite{Kim:1979if,Shifman:1979if} and 0.75 in the DFSZ-I model~\cite{Dine:1981rt,Zhitnitsky:1980tq,DiLuzio:2020wdo}. The mass itself is not predicted by the model. However, if the Peccei--Quinn symmetry is broken after inflation, simulations of the axion-string network point to $m_a\sim0.04$--$0.45\meV$~\cite{Buschmann:2021sdq,Saikawa:2024bta,Benabou:2024msj} or larger values~\cite{Gorghetto:2020qws}, so masses around a meV are motivated. They are also largely unexplored by axion haloscopes.

The difficulty at these masses is mostly geometric. In a magnetic field $\mathbf B_0$, axion DM drives a weak electric field $\mathbf E_a=-\gagg a\mathbf B_0$ that oscillates at $\nu=m_a/(2\pi)$, i.e., 242\,GHz per meV, with amplitude $E_0=\gagg a_0B_0$. Because DM is nonrelativistic, this field is nearly uniform across the depth of any device considered here. A conducting mirror placed in it must cancel the tangential field at its surface, and it does so by emitting a plane wave perpendicular to its face, with power per unit area~\cite{Horns:2012jf}
\begin{equation}
\frac{P_{\rm dish}}{A}=\frac{7.33\times10^{-26}\,{\rm W}}{\rm m^2}\left(\frac{\gagg\,{\rm GeV}}{10^{-12}}\,\frac{B_0}{10\,{\rm T}}\,\frac{\rm meV}{m_a}\right)^{\!2}\!\rho_{0.4}\,,
\label{eq:dish}
\end{equation}
where $\rho_{0.4}=\rho_a/(0.4\,{\rm GeV\,cm^{-3}})$. A resonant cavity would amplify this signal, but at 1\,meV the wavelength is $\lambda=1.24\,$mm, and a cavity resonant at this wavelength collects the signal from a tiny volume. Searches at high mass therefore keep a large converting area and enhance its emission in other ways. Dish antennas focus the emission of a bare mirror~\cite{Horns:2012jf,BREAD:2021tpx}, dielectric haloscopes place a stack of dielectric disks in front of the mirror, which shapes the photon mode so that it overlaps coherently with the axion field~\cite{Caldwell:2016dcw,Millar:2016cjp,MADMAX:2019pub}, while epsilon-near-zero (ENZ) media use a resonance of the material, where its permittivity is nearly zero~\cite{Schutte-Engel:2021bqm,Marsh:2022fmo,Mehrani:2025ryc}.

Plasma haloscopes realize the last idea with an engineered material~\cite{Lawson:2019brd,ALPHA:2022rxj}. For an electric field along the wires, a lattice of thin parallel wires responds like a plasma whose plasma frequency is set by the wire spacing, so it can be matched to the axion mass independently of the size of the device, and the axion excites the plasma oscillation coherently over the whole volume. At meV masses the problem is to get the signal out. In existing designs an antenna inside or at the edge of the medium collects the signal, which therefore has to propagate through the wire medium to reach it. At THz frequencies the wires are lossy, and the Drude response at the plasma frequency gives a field-amplitude $1/e$ attenuation length $\ell_{\rm amp}\simeq\lambda\sqrt{2Q_u}/2\pi$, consistent with the decay length $\sim\sqrt{Q}\,\lambda$ estimated in Ref.~\cite{ALPHA:2022rxj}, where $Q_u$ is the unloaded (that is to say, including only internal losses, here the ohmic dissipation in the wires) quality factor of the medium~\cite{ALPHA:2022rxj}. For a square copper-wire lattice with our choice $r/a=0.03$, using the lattice expressions of Ref.~\cite{ALPHA:2022rxj} and the surface resistance adopted below, $Q_u\simeq510\,(m_a/{\rm meV})^{-2/3}$ (about 14\% above the material quality factor $\Qm$ of Eq.~\eqref{eq:plasma} at $d=a$, because the square-lattice expression includes a correction to the inductance per wire), so that $\ell_{\rm amp}\simeq6.3\,{\rm mm}\,(m_a/{\rm meV})^{-4/3}$. A localized pickup therefore samples only a limited region of a large plasma; distributed readout requires additional collection and transmission structures.

In this Letter we let the plasma radiate. We stack a few free-standing wire grids, with the wires parallel to $\mathbf B_0$, about a quarter wavelength apart in front of a mirror (Fig.~\ref{fig:concept}). Near its resonance the stack acts as a thin ENZ slab, the axion-induced field is enhanced inside it, and the signal leaves through the whole face of the stack into the vacuum in front of it, where ordinary mirrors and lenses collect it without lateral transport to an internal pickup. The same  principle underlies the recent SQWARE~\cite{Mehrani:2025ryc} proposal, where the ENZ medium is a GaAs/AlGaAs quantum-well stack of 9--25\,cm$^2$ face tuned in a 36--50\,T field. Wire grids instead reach a large conversion area at 10\,T by repeating tiles of a conventional conductor, tuned mechanically. This converter can also search for dark photons (DPs) without a magnet. 

Geometrically, our converter is closest to a dielectric haloscope such as MADMAX~\cite{Caldwell:2016dcw,MADMAX:2019pub}, an open, mirror-backed stack of mechanically positioned layers that radiates through its face into external optics, and replaces its dielectric disks by wire grids. Like any haloscope, both shape the photon mode so that it overlaps with the spatially uniform axion field~\cite{Ioannisian:2017srr,Egge:2023cos}, but in different ways. Dielectric disks have a refractive index above unity, and in resonant configurations disks about half a wavelength apart give a field of nearly the same sign throughout, with shallow reversals in and around the disks~\cite{Millar:2016cjp}. The grids, about a quarter wavelength apart, instead act as inductive sheets, like thin layers of negative permittivity, so that just above the band edge of the lattice the stack has a refractive index close to zero and a long internal wavelength, and its field keeps the same sign throughout. End Matter B discusses the relation to earlier proposals and the role of the readout architecture.

\begin{figure}[t]
\includegraphics[width=\columnwidth]{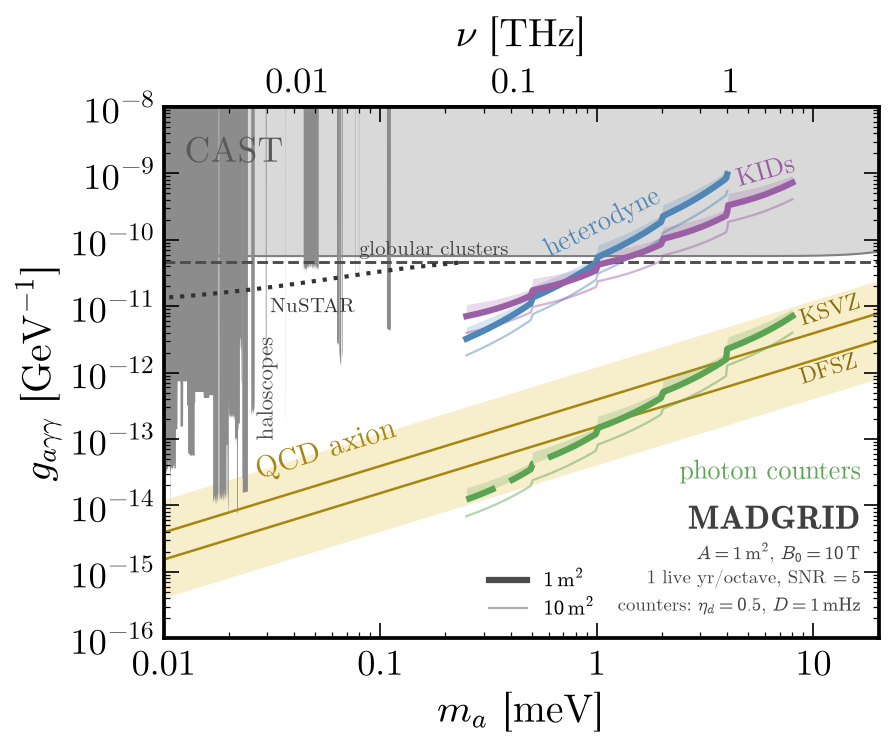}
\caption{Axion reach (local SNR\,$=5$, one live year per octave) of a 1\,m$^2$ array in 10\,T (thick curves) and of 10\,m$^2$ with the same per-channel detectors and exposure (thin curves). Photon counters assume $\eta_d=0.5$ and $D_{\rm dark}=1\,{\rm mHz}$ per channel, and are dashed below 190\,GHz, where their performance is extrapolated. Colored bands show the effect of doubling $R_s$. Existing bounds are CAST~\cite{CAST:2024eil} (light gray), globular clusters~\cite{Dolan:2022kul} (dashed), NuSTAR observations of the quiet Sun~\cite{Ruz:2024gkl} (dotted), and haloscope searches (dark gray)~\cite{Wuensch:1989sa,ADMX:2018ogs,ADMX:2021mio,CAPP:2020utb,Jeong:2020cwz,Lee:2022mnc,Yang:2023yry,Kim:2023vpo,Bae:2024kmy,Adair:2022rtw,Brubaker:2016ktl,HAYSTAC:2018rwy,HAYSTAC:2020kwv,HAYSTAC:2023cam,HAYSTAC:2024jch,TASEH:2022vvu,Grenet:2021vbb,McAllister:2017lkb,Quiskamp:2022pks,Quiskamp:2023ehr,Quiskamp:2024oet,Alesini:2019ajt,Alesini:2020vny,Alesini:2022lnp,QUAX:2023gop,QUAX:2024fut,CAST:2020rlf,Ahyoune:2024klt,MADMAX:2024sxs,GigaBREAD:2025lzq,AxionLimits}. 
}
\label{fig:reach}
\end{figure}

\begin{figure}[t]
\includegraphics[width=\columnwidth]{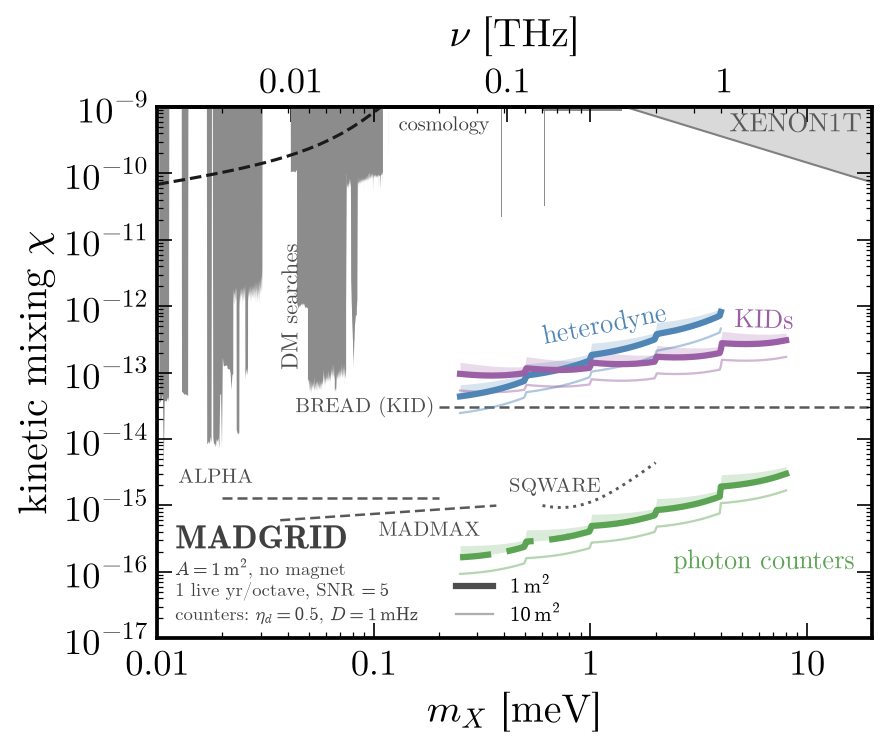}
\caption{Dark photon reach without magnet, with curves, bands and dashed extrapolations as in Fig.~\ref{fig:reach}. We show limits from solar DPs in XENON1T (light gray)~\cite{An:2020bxd}, cosmology (dashed black)~\cite{Witte:2020rvb,Caputo:2020bdy} and from direct dark-matter searches (dark gray)~\cite{Caputo:2021eaa,Dixit:2020ymh,Brun:2019kak,Ramanathan:2022egk,Schneemann:2023bqc,Bajjali:2023uis,DOSUE-RR:2022ise,DOSUE-RR:2023yml,BREAD:2023xhc,MADMAX:2024jnp,Cervantes:2022yzp,Tomita:2020usq,Knirck:2018ojz,Wei:2025ddp,Fan:2022uwu,Ourahou:2026dandelion,AxionLimits}, with the projections for a plasma haloscope and MADMAX~\cite{Gelmini:2020kcu} and for BREAD with one KID/TES over 1000 days~\cite{BREAD:2021tpx} (dashed gray), and for the most sensitive SQWARE configuration (dotted)~\cite{Mehrani:2025ryc}.}
\label{fig:dp}
\end{figure}

\textit{Mechanism.}---A grid of thin wires of radius $r$ and pitch $a$ is, for fields along the wires, a sheet of impedance $Z_g=R_g-i\omega L_g$, with $L_g=(a/2\pi)\ell$, $\ell=\ln(a/2\pi r)$~\cite{Marcuvitz:1951,Belov:2003}, and $R_g=r_sa/(2\pi r)$ from the wire resistance. Here impedances are normalized to the vacuum impedance $Z_{\rm vac}\simeq376.73\,\Omega$, and $r_s=R_s/Z_{\rm vac}$ is the dimensionless surface resistance corresponding to the physical value $R_s$ quoted below. A stack of grids at spacing $d$ therefore behaves as a Drude medium~\cite{Pendry:1996} (End Matter A), with
\begin{equation}
\varepsilon=1-\frac{\omega_p^2}{\omega(\omega+i\Gamma)},\qquad \omega_p^2=\frac{2\pi}{ad\,\ell},\qquad \Gamma=\frac{r_s}{r\ell},
\label{eq:plasma}
\end{equation}
and a material quality factor $\Qm=\omega_p/\Gamma$,
which coincides with the unloaded quality factor in the
bulk limit when wire resistance dominates the losses. Below $\omega_p$ the medium reflects like a metal, above it transmits like a dielectric, and near $\omega_p$ its permittivity is close to zero. The plasma frequency depends only on the lattice, through $ad$ and $\ell$, and not on the size of the stack, so changing the spacing $d$ tunes it. The losses enter through $\Gamma$, which is small for thick wires of a good conductor. In cold, pure copper the electron mean free path far exceeds the classical skin depth, so we use the extreme anomalous skin-effect limit, in which the surface resistance is independent of the dc resistivity and grows as $\nu^{2/3}$~\cite{ReuterSondheimer:1948}. We extrapolate the copper benchmark $R_s=2.9\,{\rm m}\Omega$, calculated at 3.9\,GHz, 4.2\,K and 6\,T~\cite{Posen:2022tbs,Calatroni:2020ase}, with this law and multiply it by 1.8 as an allowance for additional roughness and magnetoresistance losses at 10\,T, giving $R_s=83.6\,{\rm m}\Omega\,(\nu/250\,{\rm GHz})^{2/3}$. At 10\,T the cyclotron frequency of the conduction electrons, about 0.2--0.3\,THz, lies within the search range and $\mathbf B_0$ is parallel to the wire surfaces, so cyclotron effects may modify the surface impedance. The doubled-$R_s$ bands in Figs.~\ref{fig:reach} and~\ref{fig:dp} indicate the sensitivity of the forecasts to this uncertainty, which in-field THz measurements will have to settle.

Why does such a medium enhance the signal? Inside a homogeneous medium of permittivity $\varepsilon$, the
spatially uniform axion-driven field is $-E_0/\varepsilon$ rather than its vacuum value $-E_0$ (End Matter A), so near $\varepsilon=0$, this response is enhanced. However, the signal must also escape the medium to be detected. We characterize the emitted radiation by the amplitude boost factor $\beta$, defined as the outgoing electric-field amplitude divided by $E_0$, the amplitude emitted by a bare mirror. The corresponding power enhancement is $|\beta|^2$. For a slab of thickness $L$ backed by a mirror, the lossless limit $\varepsilon=\operatorname{Re}\varepsilon
\equiv u$, with $0<u\ll1$ corresponding to a frequency
slightly above the plasma frequency, admits a resonance when the slab spans half an internal wavelength, $L=\lambda/(2\sqrt{u})$, where $\lambda$ is the vacuum wavelength. At this resonance, $\beta=-2/u$ and the enhanced response inside the slab produces an enhanced outgoing wave. Equivalently, $\beta$ is the overlap of the uniform axion source with the field that a wave sent in from the detector sets up in the stack (SM~\cite{SM}). Including the leading effect of absorption, $\varepsilon\simeq u+i/\Qm$, we find the resonant power boost\footnote{This result agrees with the small-permittivity,
weak-loss limit of the mirror-backed slab response of Ref.~\cite{Mehrani:2025ryc}, after accounting
for the polarization projection used there.}
\begin{equation}
|\beta|^2\simeq
\frac{4}{[u+\pi/(2\Qm\sqrt{u})]^2}.
\label{eq:betamax}
\end{equation}
The two terms in the denominator of Eq.~\eqref{eq:betamax} are proportional, with the same factor $\pi/(2\omega\sqrt u)$, to the two rates at which the stored energy decays. In particular, $u$ corresponds to the radiative rate $\gamma_{\rm rad}$, at which energy leaves through the slab face as the signal, and $\pi/(2\Qm\sqrt u)$ to the ohmic rate $\gamma_{\rm ohm}\simeq\Gamma$, at which it is absorbed in the wires. The boost is therefore inversely proportional to the total damping, as for a driven oscillator, and the ratio of the two terms is $x\equiv\gamma_{\rm rad}/\gamma_{\rm ohm}=2\Qm u^{3/2}/\pi$ (End Matter A). A thicker slab (smaller $u$) enhances the field more but releases it more slowly, so $x$ decreases and a larger fraction of the stored energy is absorbed in the wires. The peak boost is largest, $|\beta|^2_{\max}\simeq0.61\,\Qm^{4/3}$, at $x=1/2$, i.e., when the radiative rate is half the ohmic one. For a background-dominated search, however, the thickness should optimize sensitivity across a frequency interval. We therefore choose the number of grids to maximize $\mathcal I=\int|\beta(\nu)|^4\,\dd\nu/\nu$, which accounts for signal enhancement and bandwidth (End Matter A).

The homogeneous-slab description explains the enhancement and the balance between radiation and absorption, but it assumes a small phase advance between successive grids, $kd\ll1$, with $k=\omega=2\pi/\lambda$. In the designs below $kd\simeq1.35$--2.1, so the grids are roughly a quarter of a vacuum wavelength apart and propagation across each gap must be treated explicitly. We therefore obtain the design parameters and boost curves by solving the boundary conditions for $N$ impedance sheets $Z_g$ in front of the mirror, retaining the phase accumulated between grids and their multiple reflections. To test the sheet approximation, we also solve Maxwell's equations for arrays of individual circular copper wires, including their coupling within and between grids (see the Supplemental Material (SM)~\cite{SM}).

\textit{Designs.}---For geometrically similar lattices, increasing the target axion mass requires smaller wire radii and spacings. At fixed $r/a$ and $d/a$, Eq.~\eqref{eq:plasma} gives $a\propto m_a^{-1}$. Together with the assumed copper surface resistance $R_s\propto\nu^{2/3}$, this implies $\Qm\propto m_a^{-2/3}$. The optimal slab thickness scales as $L_{\rm opt}\propto\lambda\Qm^{1/3}$, so the preferred number of grids, $N=L/d$, decreases slowly with mass, $N_{\rm opt}\propto\Qm^{1/3}$. 
To characterize the bandwidth of each design, we also define the response quality factor $\Qr=\nu_{\rm res}/\Delta\nu_{\rm FWHM}$, where $\nu_{\rm res}$ is the resonant frequency and $\Delta\nu_{\rm FWHM}$ is the full width at half maximum of the power response $|\beta(\nu)|^2$. $\Qr$ describes the resonance of the complete stack, including both radiation escaping through its open face and absorption in the wires. A larger $\Qr$ therefore corresponds to a narrower resonance. 

\begin{figure*}[t]
\includegraphics[width=0.92\textwidth]{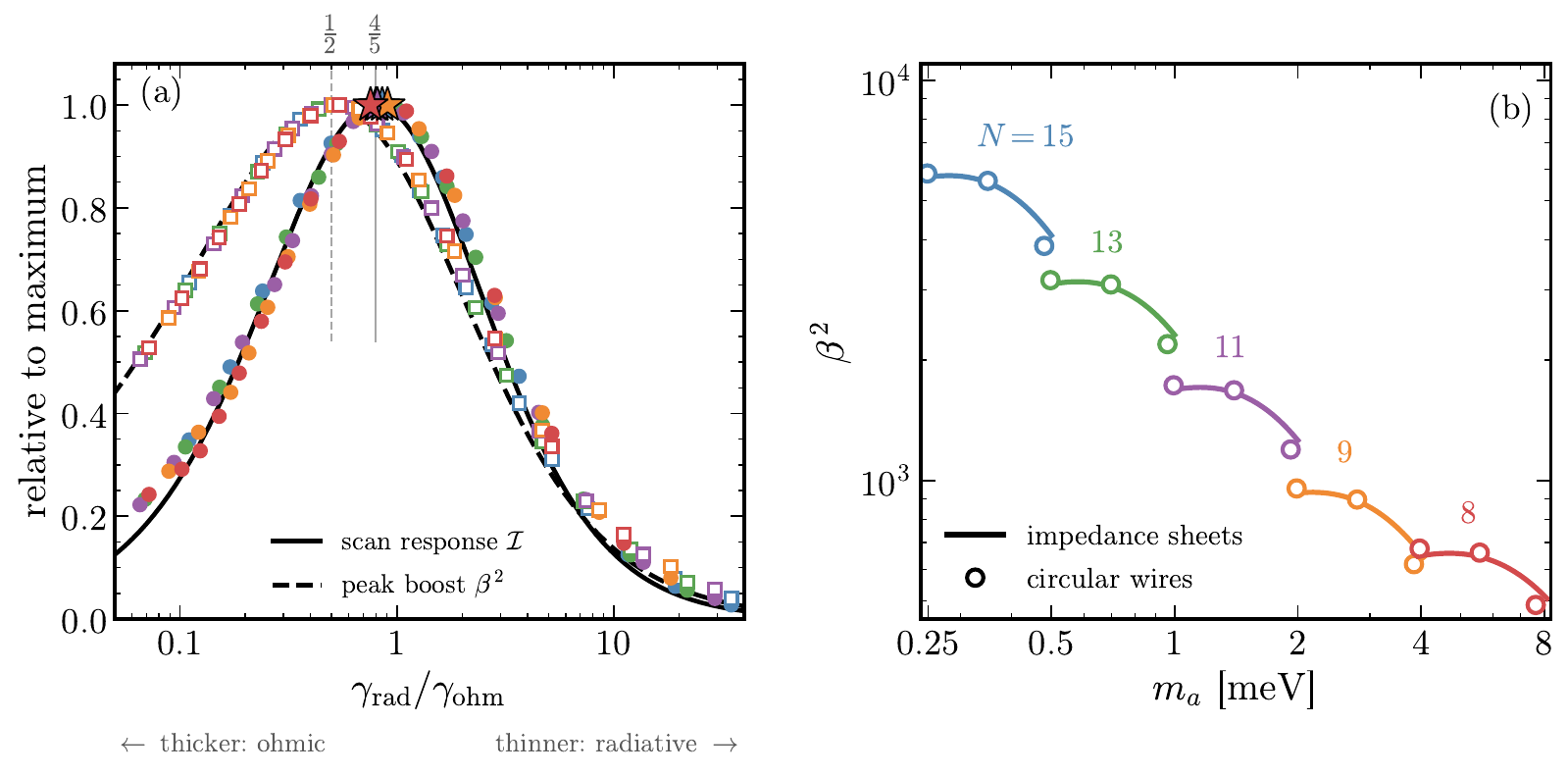}
\caption{Choice of grid number and tuning of the five cartridges of Table~\ref{tab:design}, one color per cartridge. (a) Scan response $\mathcal I$ (circles) and peak power boost (squares), normalized to their maximum within the band, for $N=3$--30 with $d=a$ and the lattice adjusted to keep the resonance at the geometric band center, plotted against $x=\gamma_{\rm rad}/\gamma_{\rm ohm}$ from the sheet calculation. Solid and dashed curves are the normalized slab predictions
and stars mark the selected designs. (b) Peak power boost as each cartridge is tuned by varying $d$ at fixed $a$, $r$ and $N$, from the sheet model (lines) and the circular-wire calculation (open circles).}
\label{fig:tuning}
\end{figure*}

Table~\ref{tab:design} lists five tunable grid stacks, or cartridges, covering successive factor-of-two intervals in axion mass. Each cartridge has a fixed wire geometry and number of grids and is tuned by varying their separation. To select the grid number, we compare designs with $N$ between 3 and 30, keeping the resonance at the center of each mass band. Figure~\ref{fig:tuning}(a) compares the normalized peak boost and scan response as functions of $x$. For $x>1$, radiation is the dominant decay channel, but for $x<1$, most of the energy is absorbed in the wires. 
The numerical results for the five cartridges agree well with the slab model.

Within each cartridge, the wire radius $r$, pitch $a$, and number of grids $N$ remain fixed. We tune the resonance by changing the common separation $d$ between successive grids, including the gap between the mirror and the first grid; compressing the stack raises the resonant frequency. The discrete-sheet calculation shows that covering each assigned interval requires reducing $d/a$ from approximately 1.69 to 0.57. Figure~\ref{fig:tuning}(b) shows the peak power boost at each tuning setting, with its maximum and minimum differing by only a factor of 1.3--1.5 within each cartridge in the sheet model. 

Tuning moves all gaps together, so the grids must remain flat and parallel and each setting stable. A common drift of 0.13--0.43\,$\mu$m lowers the on-resonance power by 10\%, so the gaps must be held or monitored at this level, with the residual drift absorbed in the scan, as cryogenic variable-delay polarization modulators already allow with sub-micron capacitive readout, although over a 1\,mm throw rather than the 16\,mm required by our lowest cartridge~\cite{Chuss:2014piper}. Sagging or tilted grids instead shift the local resonance across the aperture and degrade the coherence of the collected signal. Rms errors of $6\,\mu$m in the lowest cartridge and $0.7\,\mu$m in the highest raise the required coupling by 5\% (SM~\cite{SM}), and the degradation is gradual, so that the $5\,\mu$m already achieved over 100\,mm diameters with tungsten-wire grids~\cite{Nagler:2015xta} would raise it by at most 20\% below 2\,meV, and by factors of about 1.5--2.4 above.

\begin{table}[t]
\caption{Benchmark cartridges for five successive mass intervals. Each cartridge has a fixed wire pitch $a$, wire diameter $2r$ and grid number $N$, the latter chosen to maximize the scan response $\mathcal I$ among the sampled designs. The thickness $L=Nd$, the material and response quality factors $\Qm$ and $\Qr$, and the peak power boost $|\beta_{\rm pk}|^2$ are evaluated at the geometric center of each interval, where $d=a$, with $R_s$ taken at the resonance frequency. We use $r=0.03a$ except in the 4--8\,meV cartridge, where this scaling would require wires about $3.6\,\mu{\rm m}$ in diameter. There we take instead $2r=5\,\mu{\rm m}$, as in free-standing tungsten grids~\cite{Ade:1979,Mok:1979}, and adjust the pitch to keep the target resonance. All designs assume copper (or copper-coated) wires with the surface resistance adopted in the text.}
\label{tab:design}
\begin{ruledtabular}
\begin{tabular}{lcccccc}
Band [meV] & $N$ & $a$ [$\mu$m] & $2r$ [$\mu$m] & $L$ [mm] & $\Qm$ & $\Qr$\,/\,$|\beta_{\rm pk}|^2$\\
\hline
0.25--0.5 & 15 & 946 & 56.7 & 14.2 & 895 & 585\,/\,5530\\
0.5--1 & 13 & 474 & 28.4 & 6.16 & 564 & 379\,/\,3040\\
1--2 & 11 & 238 & 14.3 & 2.61 & 355 & 239\,/\,1650\\
2--4 & 9 & 120 & 7.17 & 1.08 & 224 & 146\,/\,883\\
4--8 & 8 & 63.7 & 5.00 & 0.510 & 169 & 123\,/\,641
\end{tabular}
\end{ruledtabular}
\end{table}

\textit{Collection.}---The preceding calculation treats the stack as finite in thickness but infinite in its plane. To determine the power reaching a receiver, we instead consider a square tile of side $W$, viewed through optics that couple to a single Gaussian spatial mode of waist radius $w\simeq W/2$, at which the field amplitude falls to $1/e$ of its central value. As in three-dimensional studies of dielectric haloscopes~\cite{MADMAX:2021lxf}, the received power is set by how well the emitted field matches this mode in amplitude and phase, rather than by the total radiated power. We compute this overlap with a model that resolves the individual wires across the tile and the finite mirror, treating the field along the wires paraxially. At the tested settings $74$--$78\%$ of the periodic-model power reaches the Gaussian mode, and an independent mode-matching calculation gives a comparable bandwidth-weighted collection factor. We adopt $\eta_{\rm ap}=0.65$, leaving some margin below these values, and describe the sampled geometries and the limitations of both calculations in the SM~\cite{SM}. Axion motion further reduces the collected signal. A velocity component parallel to the tile face produces a phase gradient in the driving current, which slightly tilts the emitted beam away from the fixed receiver mode. We include this through a factor $\xi$, the halo-averaged collected power relative to the zero-velocity result. To limit this dephasing we use 10\,cm tiles up to 4\,meV and 5\,cm tiles above, giving $\xi\simeq0.76$--$1$ across the search range for the adopted halo model, with the mean wind in the tile plane, the least favorable orientation~\cite{SM}.

\textit{Sensitivity.}---Each tile feeds its own detector, with signal power
\begin{equation}
P_s=\eta_d\eta_{\rm ap}\xi|\beta|^2P_{\rm dish},
\end{equation}
where $P_{\rm dish}$ is the power from a bare mirror of the same area. The efficiency $\eta_d$ is the product of the transmission through the receiving optics and the detector efficiency, after the collection losses already included in $\eta_{\rm ap}$ and $\xi$. We adopt $\eta_d=0.5$ as a requirement on this combined chain. For photon counting this means half of the photons entering the collected spatial mode must produce registered events. Microwave photon counters have reached comparable operational efficiencies~\cite{Balembois:2023bvs,Braggio:2024xed}, an optical efficiency of about 0.9 has been inferred for quantum-capacitance detectors at $1.5\,{\rm THz}$~\cite{Echternach:2017vtf}, and BREAD adopts the same signal efficiency~\cite{BREAD:2021tpx}.

Our reference array has an active area $A=1\,{\rm m^2}$ at $0.1\,{\rm K}$ in a field $B_0=10\,{\rm T}$, a magnet benchmark motivated by BREAD and MADMAX concepts~\cite{BREAD:2021tpx,MADMAX:2019pub}. Because the wires must be parallel to $\mathbf B_0$ while the signal leaves normal to the grids, the array requires a field parallel to the tile faces, as provided by the dipole planned for MADMAX, with the optical channels guided to detectors outside the high-field region. We allocate $T_{\rm oct}=1\,{\rm yr}$ of live observing time to each factor-of-two interval in mass, shared among tuning settings uniformly in logarithmic frequency, so that the five-octave search requires a total of five years. Sensitivities combine all tiles and settings at a local signal-to-noise ratio of 5 (End Matter B).

We consider three readouts. Heterodyne receivers resolve the axion line, of effective width $\Delta\nu_a=10^{-6}\nu$, and we assign them up to $4\,{\rm meV}$ ($0.97\,{\rm THz}$) a noise power spectral density $N_{\rm sys}=10\,\omega$, an approximate single-sideband benchmark motivated by ALMA receivers~\cite{Uzawa:2021,Kerr:2014alma,Baryshev:2015alma}. For kinetic-inductance detectors (KIDs), used as power detectors, we adopt a one-sided, absorbed-power-referred ${\rm NEP}=6\times10^{-19}\,{\rm W/\sqrt{Hz}}$ per channel, the white amplitude-readout noise of the best pixels of a kilo-pixel array near $0.85\,{\rm THz}$~\cite{Baselmans:2017kid}, although a kilo-pixel array covering $1.4$--$2.8\,{\rm THz}$ has reached a dark phase-readout NEP of about $3\times10^{-19}\,{\rm W/\sqrt{Hz}}$~\cite{Bueno:2018kid}. These demonstrations overlap the upper search band, and maintaining the same noise at lower frequencies is a development target, requiring adapted optical coupling and, below the aluminium pair-breaking threshold of about $0.39\,{\rm meV}$, a lower-gap absorber. For photon counters we assume the combined efficiency defined above and a dark-count rate $D_{\rm dark}=1\,{\rm mHz}$ per channel, as reached by quantum-dot counters~\cite{Komiyama:2000,Ikushima:2010}, giving a total background of $0.1\,{\rm s^{-1}}$ for the 100 channels of $10\,{\rm cm}$ tiles and $0.4\,{\rm s^{-1}}$ for the 400 channels of $5\,{\rm cm}$ tiles above $4\,{\rm meV}$. These forecasts assume stationary Poisson backgrounds of known mean, with cold, filtered optics keeping thermal photons below the dark rate. The stack itself absorbs nearly all the power incident in the collected mode and thus also emits thermally into it, an emission that stays below 10\% of the dark rate only if the stack is colder than about 0.4\,K at 1\,meV and 0.1\,K at the bottom of the lowest cartridge. 

Figure~\ref{fig:reach} shows that heterodyne receivers reach below the globular-cluster and CAST bounds up to approximately $1.0\,{\rm meV}$, and KIDs up to $1.2$ and $1.4\,{\rm meV}$, respectively. The assumed counters reach the DFSZ-I benchmark over approximately $0.8$--$1.2\,{\rm meV}$ and the KSVZ benchmark over $0.8$--$3.9\,{\rm meV}$, the lower endpoint being set by their assumed frequency coverage. This QCD reach defines a detector-development target whose ingredients have been demonstrated separately. Quantum-capacitance devices have counted $1.5\,{\rm THz}$ photons with high optical efficiency~\cite{Echternach:2017vtf} and have been fabricated in 441-pixel arrays~\cite{Echternach:2021}, while quantum-dot counters, which achieved the first far-infrared single-photon detection~\cite{Komiyama:2000}, have reached $1\,{\rm mHz}$ at $70\,{\rm mK}$ and 1.5--3\,THz, with efficiencies for coupling from vacuum roughly estimated at $0.1$--$1\%$~\cite{Ikushima:2010}. Further proposed readouts include among others qubit-based pair-breaking sensors~\cite{Fink:2023tvb} and Rydberg-atom photon counters with ionization readout~\cite{Banerjee:2025jss}. Combining efficient detection with millihertz backgrounds across $0.2$--$2\,{\rm THz}$ remains the main advance required. In the background-dominated regime the signal of the whole array competes with the dark counts of all its channels, so the reach of Eq.~\eqref{eq:counter_scaling} scales as $(N_{\rm ch}D_{\rm dark}/\eta_d^2)^{1/4}$, with $N_{\rm ch}=100$ for $10\,{\rm cm}$ tiles. At low masses this requirement could be relaxed by reducing the number of channels, because the axion coherence length $1/(m_a\sigma_v)\simeq0.4\,{\rm m}\,({\rm meV}/m_a)$ exceeds the tile size and neighboring tiles could be combined into a few modes, each read by one counter~\cite{Jeong:2017xqz}. 

To give an idea of how things scale, let us consider an array of $10\,{\rm cm}$ tiles at $m_a=1\,{\rm meV}$, one gets (fixed halo density, converter response, collection factors)
\begin{equation}
\begin{aligned}
g_{\min}\simeq{}&
1.30\times10^{-13}\,{\rm GeV^{-1}}
\left(\frac{10\,{\rm T}}{B_0}\right)
\left(\frac{0.5}{\eta_d}\right)^{1/2}
\\
&\times
\left(\frac{D_{\rm dark}}{1\,{\rm mHz}}\right)^{1/4}
\left(\frac{1\,{\rm m^2}}{A}\right)^{1/4}
\left(\frac{1\,{\rm yr}}{T_{\rm oct}}\right)^{1/4}.
\end{aligned}
\label{eq:counter_scaling}
\end{equation}


\textit{Dark photons.}---The same converter can search for DP DM without an applied magnetic field. A DP of mass $m_X$ and dimensionless kinetic mixing $\chi$~\cite{Holdom:1985ag} drives ordinary electric currents through an effective source $\mathbf J_X=\chi m_X^2\mathbf X$. For local density $\rho_X$, its driving-field amplitude along the wires is $\chi\sqrt{2\rho_X}\cos\theta$, where $\theta$ is the angle between the DP polarization and the wires~\cite{Gelmini:2020kcu,Caputo:2021eaa}. Replacing the axion amplitude $\gagg a_0B_0$ by this expression gives 
\begin{equation}
\begin{aligned}
\chi_{\min}\simeq{}&
3.38\times10^{-15}
\left(\frac{g_{\min}}{10^{-12}\,{\rm GeV^{-1}}}\right)
\left(\frac{B_{\rm ref}}{10\,{\rm T}}\right)
\\
&\times
\left(\frac{1\,{\rm meV}}{m_X}\right)
\left(\frac{1/3}{\kappa}\right)^{1/2},
\end{aligned}
\label{eq:dpmap}
\end{equation}
where $B_{\rm ref}$ is the field used in the axion forecast and $\kappa=\langle\cos^2\theta\rangle=1/3$ for isotropically randomized polarization. A fixed polarization produces daily modulation as the wires rotate with the Earth, making the sensitivity depend on detector orientation and observing schedule~\cite{Caputo:2021eaa}. We assume randomized polarization and show our results in Figure~\ref{fig:dp}. 

\textit{Conclusions.}---With KIDs at ${\rm NEP}=6\times10^{-19}\,{\rm W/\sqrt{Hz}}$, a 1\,m$^2$ MADGRID array would set the strongest constraints on axions below about 1.2--1.4\,meV and on DPs near 1\,meV. The next steps are a bench demonstration of the tile resonance and its spacing tuning, and a measurement of the in-field THz surface impedance of thin copper wires. THz photon counters with millihertz backgrounds would then bring the QCD axion band within reach.

\textit{Acknowledgments---}
It is a pleasure to thank Valerie Domcke, Sebastian Ellis, Anton Ivanov, Bela Majorovits and Georg Raffelt for discussions and comments on the draft. This work is supported by an ERC STG grant (``AstroDarkLS'', grant No. 101117510). The author acknowledges hospitality by the Institute for Advanced Study supported by an anonymous donor.

\bibliography{bib}

@article{DiLuzio:2020wdo,
    author = "Di Luzio, Luca and Giannotti, Maurizio and Nardi, Enrico and Visinelli, Luca",
    title = "{The landscape of QCD axion models}",
    reportNumber = "DESY 20-036, DESY-20-036",
    doi = "10.1016/j.physrep.2020.06.002",
    journal = "Phys. Rept.",
    volume = "870",
    pages = "1--117",
    year = "2020"
}

@article{Buschmann:2021sdq,
    author = "Buschmann, Malte and Foster, Joshua W. and Hook, Anson and Peterson, Adam and Willcox, Don E. and Zhang, Weiqun and Safdi, Benjamin R.",
    title = "{Dark matter from axion strings with adaptive mesh refinement}",
    doi = "10.1038/s41467-022-28669-y",
    journal = "Nature Commun.",
    volume = "13",
    number = "1",
    pages = "1049",
    year = "2022"
}

@article{Gorghetto:2020qws,
    author = "Gorghetto, Marco and Hardy, Edward and Villadoro, Giovanni",
    title = "{More axions from strings}",
    doi = "10.21468/SciPostPhys.10.2.050",
    journal = "SciPost Phys.",
    volume = "10",
    number = "2",
    pages = "050",
    year = "2021"
}

@article{Saikawa:2024bta,
    author = "Saikawa, Ken'ichi and Redondo, Javier and Vaquero, Alejandro and Kaltschmidt, Mathieu",
    title = "{Spectrum of global string networks and the axion dark matter mass}",
    reportNumber = "KANAZAWA-24-02, MPP-2024-18",
    doi = "10.1088/1475-7516/2024/10/043",
    journal = "JCAP",
    volume = "2024",
    number = "10",
    pages = "043",
    year = "2024"
}

@article{HAYSTAC:2020kwv,
    author = "Backes, K. M. and others",
    collaboration = "HAYSTAC",
    title = "{A quantum-enhanced search for dark matter axions}",
    doi = "10.1038/s41586-021-03226-7",
    journal = "Nature",
    volume = "590",
    number = "7845",
    pages = "238--242",
    year = "2021"
}

@article{Quiskamp:2022pks,
    author = "Quiskamp, Aaron P. and McAllister, Ben T. and Altin, Paul and Ivanov, Eugene N. and Goryachev, Maxim and Tobar, Michael E.",
    collaboration = "ORGAN",
    title = "{Direct search for dark matter axions excluding ALP cogenesis in the 63- to 67-$\mu$eV range with the ORGAN experiment}",
    eprint = "2203.12152",
    archivePrefix = "arXiv",
    primaryClass = "hep-ex",
    doi = "10.1126/sciadv.abq3765",
    journal = "Sci. Adv.",
    volume = "8",
    number = "27",
    pages = "eabq3765",
    year = "2022"
}

@article{CAST:2024eil,
    author = {Altenm{\"u}ller, K. and others},
    collaboration = "CAST",
    title = "{New Upper Limit on the Axion-Photon Coupling with an Extended CAST Run with a Xe-Based Micromegas Detector}",
    doi = "10.1103/PhysRevLett.133.221005",
    journal = "Phys. Rev. Lett.",
    volume = "133",
    number = "22",
    pages = "221005",
    year = "2024"
}

@article{Dolan:2022kul,
    author = "Dolan, Matthew J. and Hiskens, Frederick J. and Volkas, Raymond R.",
    title = "{Advancing globular cluster constraints on the axion-photon coupling}",
    doi = "10.1088/1475-7516/2022/10/096",
    journal = "JCAP",
    volume = "2022",
    number = "10",
    pages = "096",
    year = "2022"
}

@article{Horns:2012jf,
    author = "Horns, Dieter and Jaeckel, Joerg and Lindner, Axel and Lobanov, Andrei and Redondo, Javier and Ringwald, Andreas",
    title = "{Searching for WISPy Cold Dark Matter with a Dish Antenna}",
    reportNumber = "DESY-12-227, MPP-2012-158",
    doi = "10.1088/1475-7516/2013/04/016",
    journal = "JCAP",
    volume = "2013",
    number = "04",
    pages = "016",
    year = "2013"
}

@article{Caldwell:2016dcw,
    author = "Caldwell, Allen and Dvali, Gia and Majorovits, B{\'e}la and Millar, Alexander and Raffelt, Georg and Redondo, Javier and Reimann, Olaf and Simon, Frank and Steffen, Frank",
    collaboration = "MADMAX Working Group",
    title = "{Dielectric Haloscopes: A New Way to Detect Axion Dark Matter}",
    doi = "10.1103/PhysRevLett.118.091801",
    journal = "Phys. Rev. Lett.",
    volume = "118",
    number = "9",
    pages = "091801",
    year = "2017"
}

@article{Millar:2016cjp,
    author = "Millar, Alexander J. and Raffelt, Georg G. and Redondo, Javier and Steffen, Frank D.",
    title = "{Dielectric Haloscopes to Search for Axion Dark Matter: Theoretical Foundations}",
    doi = "10.1088/1475-7516/2017/01/061",
    journal = "JCAP",
    volume = "2017",
    number = "01",
    pages = "061",
    year = "2017"
}

@article{MADMAX:2019pub,
    author = "Brun, P. and others",
    collaboration = "MADMAX",
    title = "{A new experimental approach to probe QCD axion dark matter in the mass range above 40 $\mu$eV}",
    reportNumber = "DESY-19-011, DESY 19-011",
    doi = "10.1140/epjc/s10052-019-6683-x",
    journal = "Eur. Phys. J. C",
    volume = "79",
    number = "3",
    pages = "186",
    year = "2019"
}

@article{Lawson:2019brd,
    author = "Lawson, Matthew and Millar, Alexander J. and Pancaldi, Matteo and Vitagliano, Edoardo and Wilczek, Frank",
    title = "{Tunable axion plasma haloscopes}",
    reportNumber = "NORDITA-2019-038, MIT-CTP-5116, MPP-2019-83",
    doi = "10.1103/PhysRevLett.123.141802",
    journal = "Phys. Rev. Lett.",
    volume = "123",
    number = "14",
    pages = "141802",
    year = "2019"
}

@article{ALPHA:2022rxj,
    author = "Millar, Alexander J. and others",
    collaboration = "ALPHA",
    title = "{Searching for dark matter with plasma haloscopes}",
    reportNumber = "FERMILAB-PUB-22-739-T",
    doi = "10.1103/PhysRevD.107.055013",
    journal = "Phys. Rev. D",
    volume = "107",
    number = "5",
    pages = "055013",
    year = "2023"
}

@article{BREAD:2021tpx,
    author = "Liu, Jesse and others",
    collaboration = "BREAD",
    title = "{Broadband Solenoidal Haloscope for Terahertz Axion Detection}",
    reportNumber = "FERMILAB-PUB-21-694-AD-PPD-TD",
    doi = "10.1103/PhysRevLett.128.131801",
    journal = "Phys. Rev. Lett.",
    volume = "128",
    number = "13",
    pages = "131801",
    year = "2022"
}

@article{Mehrani:2025ryc,
    author = "Mehrani, Jaanita and Xu, Tao and Baydin, Andrey and Manfra, Michael J. and Everitt, Henry O. and Long, Andrew J. and Sinha, Kuver and Kono, Junichiro and Huang, Shengxi",
    title = "{Quantum Semiconductor Heterostructures for meV Axion Dark Matter Detection}",
    doi = "10.1103/y7jl-gj2k",
    journal = "Phys. Rev. Lett.",
    volume = "136",
    number = "24",
    pages = "241001",
    year = "2026"
}

@article{Schutte-Engel:2021bqm,
    author = {Sch{\"u}tte-Engel, Jan and Marsh, David J. E. and Millar, Alexander J. and Sekine, Akihiko and Chadha-Day, Francesca and Hoof, Sebastian and Ali, Mazhar N. and Fong, Kin-Chung and Hardy, Edward and {\v{S}}mejkal, Libor},
    title = "{Axion quasiparticles for axion dark matter detection}",
    reportNumber = "IPPP/20/78, NORDITA-2021-007",
    doi = "10.1088/1475-7516/2021/08/066",
    journal = "JCAP",
    volume = "2021",
    number = "08",
    pages = "066",
    year = "2021"
}

@article{Jeong:2017xqz,
    author = "Jeong, Junu and Youn, SungWoo and Ahn, Saebyeok and Kang, Chanshin and Semertzidis, Yannis K.",
    title = "{Phase-matching of multiple-cavity detectors for dark matter axion search}",
    eprint = "1707.05925",
    archivePrefix = "arXiv",
    primaryClass = "astro-ph.IM",
    doi = "10.1016/j.astropartphys.2017.10.012",
    journal = "Astropart. Phys.",
    volume = "97",
    pages = "33--37",
    year = "2018"
}

@article{Balembois:2023bvs,
    author = "Balembois, L{\'e}o and Travesedo, Jaime and Pallegoix, Louis and May, Alexandre and Billaud, Eric and Villiers, Marius and Est{\`e}ve, Daniel and Vion, Denis and Bertet, Patrice and Flurin, Emmanuel",
    title = "{Cyclically Operated Microwave Single-Photon Counter with Sensitivity of $10^{-22}~\mathrm{W}/\sqrt{\mathrm{Hz}}$}",
    eprint = "2307.03614",
    archivePrefix = "arXiv",
    primaryClass = "quant-ph",
    doi = "10.1103/PhysRevApplied.21.014043",
    journal = "Phys. Rev. Applied",
    volume = "21",
    number = "1",
    pages = "014043",
    year = "2024"
}

@article{Braggio:2024xed,
    author = "Braggio, C. and others",
    title = "{Quantum-Enhanced Sensing of Axion Dark Matter with a Transmon-Based Single Microwave Photon Counter}",
    reportNumber = "FERMILAB-PUB-24-0510-SQMS-V",
    doi = "10.1103/PhysRevX.15.021031",
    journal = "Phys. Rev. X",
    volume = "15",
    number = "2",
    pages = "021031",
    year = "2025"
}

@article{Dixit:2020ymh,
    author = "Dixit, Akash V. and Chakram, Srivatsan and He, Kevin and Agrawal, Ankur and Naik, Ravi K. and Schuster, David I. and Chou, Aaron",
    title = "{Searching for Dark Matter with a Superconducting Qubit}",
    reportNumber = "FERMILAB-PUB-20-424-E",
    doi = "10.1103/PhysRevLett.126.141302",
    journal = "Phys. Rev. Lett.",
    volume = "126",
    number = "14",
    pages = "141302",
    year = "2021"
}

@article{Nagler:2015xta,
    author = "Nagler, Peter C. and Fixsen, Dale J. and Kogut, Alan and Tucker, Gregory S.",
    title = "{Systematic effects in polarizing Fourier transform spectrometers for cosmic microwave background observations}",
    eprint = "1510.08089",
    archivePrefix = "arXiv",
    primaryClass = "astro-ph.IM",
    doi = "10.1088/0067-0049/221/1/21",
    journal = "Astrophys. J. Suppl.",
    volume = "221",
    number = "1",
    pages = "21",
    year = "2015"
}

@article{Posen:2022tbs,
    author = "Posen, S. and Checchin, M. and Melnychuk, O. S. and Ring, T. and Gonin, I. and Khabiboulline, T.",
    title = "{High-Quality-Factor Superconducting Cavities in Tesla-Scale Magnetic Fields for Dark-Matter Searches}",
    reportNumber = "FERMILAB-PUB-22-066-SQMS-TD",
    doi = "10.1103/PhysRevApplied.20.034004",
    journal = "Phys. Rev. Applied",
    volume = "20",
    number = "3",
    pages = "034004",
    year = "2023"
}

@article{Peccei:1977hh,
    author = "Peccei, R. D. and Quinn, Helen R.",
    title = "{CP Conservation in the Presence of Pseudoparticles}",
    reportNumber = "ITP-568-STANFORD",
    doi = "10.1103/PhysRevLett.38.1440",
    journal = "Phys. Rev. Lett.",
    volume = "38",
    pages = "1440--1443",
    year = "1977"
}

@article{Weinberg:1977ma,
    author = "Weinberg, Steven",
    title = "{A New Light Boson?}",
    reportNumber = "HUTP-77/A074",
    doi = "10.1103/PhysRevLett.40.223",
    journal = "Phys. Rev. Lett.",
    volume = "40",
    pages = "223--226",
    year = "1978"
}

@article{Wilczek:1977pj,
    author = "Wilczek, Frank",
    title = "{Problem of Strong  $P$  and  $T$  Invariance in the Presence of Instantons}",
    reportNumber = "Print-77-0939 (COLUMBIA)",
    doi = "10.1103/PhysRevLett.40.279",
    journal = "Phys. Rev. Lett.",
    volume = "40",
    pages = "279--282",
    year = "1978"
}

@article{Kim:1979if,
    author = "Kim, Jihn E.",
    title = "{Weak Interaction Singlet and Strong CP Invariance}",
    reportNumber = "UPR-0120T",
    doi = "10.1103/PhysRevLett.43.103",
    journal = "Phys. Rev. Lett.",
    volume = "43",
    pages = "103",
    year = "1979"
}

@article{Shifman:1979if,
    author = "Shifman, Mikhail A. and Vainshtein, A. I. and Zakharov, Valentin I.",
    title = "{Can Confinement Ensure Natural CP Invariance of Strong Interactions?}",
    reportNumber = "ITEP-64-1979",
    doi = "10.1016/0550-3213(80)90209-6",
    journal = "Nucl. Phys. B",
    volume = "166",
    pages = "493--506",
    year = "1980"
}

@article{Dine:1981rt,
    author = "Dine, Michael and Fischler, Willy and Srednicki, Mark",
    title = "{A Simple Solution to the Strong CP Problem with a Harmless Axion}",
    reportNumber = "Print-81-0320 (IAS,PRINCETON)",
    doi = "10.1016/0370-2693(81)90590-6",
    journal = "Phys. Lett. B",
    volume = "104",
    pages = "199--202",
    year = "1981"
}

@article{Zhitnitsky:1980tq,
    author = "Zhitnitsky, A. R.",
    title = "{On Possible Suppression of the Axion Hadron Interactions. (In Russian)}",
    journal = "Sov. J. Nucl. Phys.",
    volume = "31",
    pages = "260",
    year = "1980"
}

@article{Preskill:1982cy,
    author = "Preskill, John and Wise, Mark B. and Wilczek, Frank",
    editor = "Srednicki, M. A.",
    title = "{Cosmology of the Invisible Axion}",
    reportNumber = "HUTP-82-A048, NSF-ITP-82-103",
    doi = "10.1016/0370-2693(83)90637-8",
    journal = "Phys. Lett. B",
    volume = "120",
    pages = "127--132",
    year = "1983"
}

@article{Abbott:1982af,
    author = "Abbott, L. F. and Sikivie, P.",
    editor = "Srednicki, M. A.",
    title = "{A Cosmological Bound on the Invisible Axion}",
    reportNumber = "PRINT-82-0695 (BRANDEIS)",
    doi = "10.1016/0370-2693(83)90638-X",
    journal = "Phys. Lett. B",
    volume = "120",
    pages = "133--136",
    year = "1983"
}

@article{Dine:1982ah,
    author = "Dine, Michael and Fischler, Willy",
    editor = "Srednicki, M. A.",
    title = "{The Not So Harmless Axion}",
    reportNumber = "UPR-0201T",
    doi = "10.1016/0370-2693(83)90639-1",
    journal = "Phys. Lett. B",
    volume = "120",
    pages = "137--141",
    year = "1983"
}

@article{Echternach:2017vtf,
    author = "Echternach, P. M. and Pepper, B. J. and Reck, T. and Bradford, C. M.",
    title = "{Single photon detection of 1.5{\,}THz radiation with the quantum capacitance detector}",
    doi = "10.1038/s41550-017-0294-y",
    journal = "Nature Astron.",
    volume = "2",
    number = "1",
    pages = "90--97",
    year = "2018"
}

@article{Pendry:1996,
    author = "Pendry, J. B. and Holden, A. J. and Stewart, W. J. and Youngs, I.",
    title = "{Extremely Low Frequency Plasmons in Metallic Mesostructures}",
    journal = "Phys. Rev. Lett.",
    volume = "76",
    pages = "4773--4776",
    year = "1996",
    doi = "10.1103/PhysRevLett.76.4773"
}

@article{Belov:2003,
    author = "Belov, P. A. and Marqu\'es, R. and Maslovski, S. I. and Nefedov, I. S. and Silveirinha, M. and Simovski, C. R. and Tretyakov, S. A.",
    title = "{Strong spatial dispersion in wire media in the very large wavelength limit}",
    journal = "Phys. Rev. B",
    volume = "67",
    pages = "113103",
    year = "2003",
    doi = "10.1103/PhysRevB.67.113103"
}

@article{ReuterSondheimer:1948,
    author = "Reuter, G. E. H. and Sondheimer, E. H.",
    title = "{The theory of the anomalous skin effect in metals}",
    journal = "Proc. R. Soc. A",
    volume = "195",
    pages = "336--364",
    year = "1948",
    doi = "10.1098/rspa.1948.0123"
}

@book{Marcuvitz:1951,
    author = "Marcuvitz, N.",
    title = "{Waveguide Handbook}",
    publisher = "McGraw-Hill",
    address = "New York",
    year = "1951"
}

@misc{AxionLimits,
    author = "O'Hare, C.",
    title = "{cajohare/AxionLimits: AxionLimits}",
    howpublished = "\url{https://cajohare.github.io/AxionLimits/}",
    year = "2020",
    doi = "10.5281/zenodo.3932430"
}

@article{Chuss:2014piper,
    author = "Chuss, D. T. and Eimer, J. R. and Fixsen, D. J. and Hinderks, J. and Kogut, A. J. and Lazear, J. and Mirel, P. and Switzer, E. and Voellmer, G. M. and Wollack, E. J.",
    title = "{Variable-delay polarization modulators for cryogenic millimeter-wave applications}",
    journal = "Rev. Sci. Instrum.",
    volume = "85",
    pages = "064501",
    year = "2014",
    doi = "10.1063/1.4879499",
}

@article{Quiskamp:2023ehr,
    author = "Quiskamp, Aaron and McAllister, Ben T. and Altin, Paul and Ivanov, Eugene N. and Goryachev, Maxim and Tobar, Michael E.",
    title = "{Exclusion of Axionlike-Particle Cogenesis Dark Matter in a Mass Window above 100{\,}{\,}{\ensuremath{\mu}}eV}",
    doi = "10.1103/PhysRevLett.132.031601",
    journal = "Phys. Rev. Lett.",
    volume = "132",
    number = "3",
    pages = "031601",
    year = "2024"
}

@article{Fink:2023tvb,
    author = "Fink, Caleb W. and Salemi, Chiara P. and Young, Betty A. and Schuster, David I. and Kurinsky, Noah A.",
    title = "{The Superconducting Quasiparticle-Amplifying Transmon: A Qubit-Based Sensor for meV Scale Phonons and Single THz Photons}",
    eprint = "2310.01345",
    archivePrefix = "arXiv",
    primaryClass = "physics.ins-det",
    reportNumber = "LA-UR-23-30997",
    doi = "10.1103/PhysRevApplied.22.054009",
    journal = "Phys. Rev. Applied",
    volume = "22",
    number = "5",
    pages = "054009",
    year = "2024"
}

@article{Ruz:2024gkl,
    author = "Ruz, J. and others",
    title = "{NuSTAR as an Axion Helioscope}",
    doi = "10.1103/18sn-hxtb",
    journal = "Phys. Rev. Lett.",
    volume = "135",
    number = "14",
    pages = "141001",
    year = "2025"
}

@article{Echternach:2021,
  author={Echternach, P. M. and Beyer, A. D. and Bradford, C. M.},
  title={{Large array of low-frequency readout quantum capacitance detectors}},
  journal={J. Astron. Telesc. Instrum. Syst.}, volume={7}, pages={011003}, year={2021}, doi={10.1117/1.JATIS.7.1.011003}}

@article{Uzawa:2021,
  author={Uzawa, Y. and Fujii, Y. and Kojima, T. and Kroug, M. and Shan, W. and Ezaki, S. and Miyachi, A. and Kiuchi, H. and Gonzalez, A.},
  title={{Superconducting Receiver Technologies Supporting ALMA and Future Prospects}},
  journal={Radio Sci.}, volume={56}, pages={e2020RS007157}, year={2021}, doi={10.1029/2020RS007157}}

@article{Ade:1979,
  author={Ade, P. A. R. and Costley, A. E. and Cunningham, C. T. and Mok, C. L. and Neill, G. F. and Parker, T. J.},
  title={{Free-standing grids wound from 5 $\mu$m diameter wire for spectroscopy at far-infrared wavelengths}},
  journal={Infrared Phys.}, volume={19}, pages={599}, year={1979}, doi={10.1016/0020-0891(79)90080-0}}

@article{Egge:2023cos,
    author = "Egge, J. and others",
    title = "{Experimental determination of axion signal power of dish antennas and dielectric haloscopes using the reciprocity approach}",
    reportNumber = "FERMILAB-PUB-24-0399-PPD",
    doi = "10.1088/1475-7516/2024/04/005",
    journal = "JCAP",
    volume = "2024",
    number = "04",
    pages = "005",
    year = "2024"
}

@techreport{Calatroni:2020ase,
    author = "Calatroni, Sergio",
    title = "{A Mathematica Notebook for the calculation of the anomalous skin effect in copper}",
    institution = "CERN",
    address = "Geneva",
    number = "CERN-ACC-NOTE-2020-0029",
    year = "2020",
    doi = "10.17181/CERN.MZIG.TJNX",
    url = "https://cds.cern.ch/record/2718002"
}

@article{MADMAX:2024sxs,
    author = "{Ary dos Santos Garcia}, B. and others",
    collaboration = "MADMAX",
    title = "{First Search for Axion Dark Matter with a MADMAX Prototype}",
    reportNumber = "FERMILAB-PUB-24-0095-PPD",
    doi = "10.1103/c749-419q",
    journal = "Phys. Rev. Lett.",
    volume = "135",
    number = "4",
    pages = "041001",
    year = "2025"
}

@article{Marsh:2022fmo,
    author = {Marsh, David J. E. and McDonald, Jamie I. and Millar, Alexander J. and Sch{\"u}tte-Engel, Jan},
    title = "{Axion detection with phonon-polaritons revisited}",
    reportNumber = "KCL-PH-TH/2022-51, CP3-22-46, FERMILAB-PUB-22-736-T",
    doi = "10.1103/PhysRevD.107.035036",
    journal = "Phys. Rev. D",
    volume = "107",
    number = "3",
    pages = "035036",
    year = "2023"
}

@article{MADMAX:2021lxf,
    author = "Knirck, S. and others",
    collaboration = "MADMAX",
    title = "{Simulating MADMAX in 3D: requirements for dielectric axion haloscopes}",
    reportNumber = "MPP-2021-61, FERMILAB-PUB-22-012-PPD, DESY-21-051",
    doi = "10.1088/1475-7516/2021/10/034",
    journal = "JCAP",
    volume = "2021",
    number = "10",
    pages = "034",
    year = "2021"
}

@article{Egge:2022gfp,
    author = "Egge, Jacob",
    title = "{Axion haloscope signal power from reciprocity}",
    doi = "10.1088/1475-7516/2023/04/064",
    journal = "JCAP",
    volume = "2023",
    number = "04",
    pages = "064",
    year = "2023"
}

@article{DeMiguel:2020rpn,
    author = "De Miguel, Javier",
    title = "{A dark matter telescope probing the 6 to 60 GHz band}",
    doi = "10.1088/1475-7516/2021/04/075",
    journal = "JCAP",
    volume = "2021",
    number = "04",
    pages = "075",
    year = "2021"
}

@article{GigaBREAD:2025lzq,
    author = "Hoshino, Gabe and others",
    collaboration = "GigaBREAD",
    title = "{First Axionlike Particle Results from a Broadband Search for Wavelike Dark Matter in the 44 to 52{\,}{\,}{\ensuremath{\mu}}eV Range with a Coaxial Dish Antenna}",
    reportNumber = "FERMILAB-PUB-25-0076-PPD",
    doi = "10.1103/PhysRevLett.134.171002",
    journal = "Phys. Rev. Lett.",
    volume = "134",
    number = "17",
    pages = "171002",
    year = "2025"
}

@article{BREAD:2023xhc,
    author = "Knirck, Stefan and others",
    collaboration = "BREAD",
    title = "{First Results from a Broadband Search for Dark Photon Dark Matter in the 44 to 52{\,}{\,}{\ensuremath{\mu}}eV Range with a Coaxial Dish Antenna}",
    reportNumber = "FERMILAB-PUB-23-625-PPD",
    doi = "10.1103/PhysRevLett.132.131004",
    journal = "Phys. Rev. Lett.",
    volume = "132",
    number = "13",
    pages = "131004",
    year = "2024"
}

@article{Banerjee:2025jss,
    author = "Banerjee, Abhishek and Ebadi, Reza and Rajendran, Surjeet",
    title = "{Rydberg single photon detection for probing 0.1-10 meV dark matter with BREAD}",
    doi = "10.1103/yj2s-v785",
    journal = "Phys. Rev. D",
    volume = "114",
    number = "5",
    pages = "055011",
    year = "2026"
}

@article{Quiskamp:2024oet,
    author = "Quiskamp, Aaron P. and Flower, Graeme R. and Samuels, Steven and McAllister, Ben T. and Altin, Paul and Ivanov, Eugene N. and Goryachev, Maxim and Tobar, Michael E.",
    collaboration = "ORGAN",
    title = "{Near-quantum-limited axion dark matter search with the ORGAN experiment around 26{\,}{\,}{\ensuremath{\mu}}eV}",
    doi = "10.1103/PhysRevD.111.095007",
    journal = "Phys. Rev. D",
    volume = "111",
    number = "9",
    pages = "095007",
    year = "2025"
}

@article{QUAX:2024fut,
    author = "Rettaroli, A. and others",
    collaboration = "QUAX",
    title = "{Search for axion dark matter with the QUAX{\textendash}LNF tunable haloscope}",
    reportNumber = "FERMILAB-PUB-24-0511-SQMS-V",
    doi = "10.1103/PhysRevD.110.022008",
    journal = "Phys. Rev. D",
    volume = "110",
    number = "2",
    pages = "022008",
    year = "2024"
}

@article{Kerr:2014alma,
    author = "Kerr, Anthony R. and Pan, Shing-Kuo and Claude, Stephane M. X. and Dindo, Philip and Lichtenberger, Arthur W. and Effland, John E. and Lauria, Eugene F.",
    title = "{Development of the ALMA Band-3 and Band-6 Sideband-Separating SIS Mixers}",
    journal = "IEEE Trans. Terahertz Sci. Technol.",
    volume = "4",
    number = "2",
    pages = "201--212",
    year = "2014",
    doi = "10.1109/TTHZ.2014.2302537"
}

@article{Baryshev:2015alma,
    author = "Baryshev, A. M. and Hesper, R. and Mena, F. P. and Klapwijk, T. M. and van Kempen, T. A. and Hogerheijde, M. R. and others",
    title = "{The ALMA Band 9 receiver: Design, construction, characterization, and first light}",
    journal = "Astron. Astrophys.",
    volume = "577",
    pages = "A129",
    year = "2015",
    doi = "10.1051/0004-6361/201425529"
}

@article{Ikushima:2010,
    author = "Ikushima, Kenji and Komiyama, Susumu",
    title = "{Imaging by terahertz photon counting}",
    journal = "C. R. Physique",
    volume = "11",
    number = "7-8",
    pages = "444--456",
    year = "2010",
    doi = "10.1016/j.crhy.2010.06.009"
}

@article{Komiyama:2000,
    author = "Komiyama, S. and Astafiev, O. and Antonov, V. and Kutsuwa, T. and Hirai, H.",
    title = "{A single-photon detector in the far-infrared range}",
    journal = "Nature",
    volume = "403",
    pages = "405--407",
    year = "2000",
    doi = "10.1038/35000166"
}

@article{Bueno:2018kid,
  author={Bueno, J. and Murugesan, V. and Karatsu, K. and Thoen, D. J. and Baselmans, J. J. A.},
  title={Ultrasensitive kilo-pixel imaging array of photon noise-limited kinetic inductance detectors over an octave of bandwidth for {THz} astronomy},
  journal={J. Low Temp. Phys.}, volume={193}, pages={96}, year={2018}, doi={10.1007/s10909-018-1962-8}}

@article{Baselmans:2017kid,
  author={Baselmans, J. J. A. and others},
  title={A kilo-pixel imaging system for future space based far-infrared observatories using microwave kinetic inductance detectors},
  journal={Astron. Astrophys.}, volume={601}, pages={A89}, year={2017}, doi={10.1051/0004-6361/201629653}}

@article{Gelmini:2020kcu,
  author={Gelmini, Graciela B. and Millar, Alexander J. and Takhistov, Volodymyr and Vitagliano, Edoardo},
  title={Probing dark photons with plasma haloscopes},
  journal={Phys. Rev. D}, volume={102}, pages={043003}, year={2020}, doi={10.1103/PhysRevD.102.043003}, eprint={2006.06836}, archivePrefix={arXiv}}

@article{Caputo:2021eaa,
  author={Caputo, Andrea and Millar, Alexander J. and O'Hare, Ciaran A. J. and Vitagliano, Edoardo},
  title={Dark photon limits: A handbook},
  journal={Phys. Rev. D}, volume={104}, pages={095029}, year={2021}, doi={10.1103/PhysRevD.104.095029}, eprint={2105.04565}, archivePrefix={arXiv}}

@article{Holdom:1985ag,
  author={Holdom, Bob},
  title={{Two U(1)'s and $\epsilon$ charge shifts}},
  journal={Phys. Lett. B}, volume={166}, pages={196}, year={1986}, doi={10.1016/0370-2693(86)91377-8}}

@article{Knirck:2018ojz,
  author={Knirck, Stefan and Yamazaki, Takayuki and Okesaku, Yoshiki and Asai, Shoji and Idehara, Toshitaka and Inada, Toshiaki},
  title={{First results from a hidden photon dark matter search in the meV sector using a plane-parabolic mirror system}},
  journal={JCAP}, volume = "2018",
    number = "11", pages={031}, year={2018}, doi={10.1088/1475-7516/2018/11/031}, eprint={1806.05120}, archivePrefix={arXiv}}

@article{Fan:2022uwu,
  author={Fan, Xing and Gabrielse, Gerald and Graham, Peter W. and Harnik, Roni and Myers, Thomas G. and Ramani, Harikrishnan and others},
  title={{One-Electron Quantum Cyclotron as a Milli-eV Dark-Photon Detector}},
  journal={Phys. Rev. Lett.}, volume={129}, pages={261801}, year={2022}, doi={10.1103/PhysRevLett.129.261801}, eprint={2208.06519}, archivePrefix={arXiv}}

@article{Wei:2025ddp,
  author={Wei, Guoqing and Wu, Diguang and Kang, Runqi and Jiang, Qingning and Jiao, Man and Rong, Xing and others},
  title={{New Constraints on Dark Photon Dark Matter with a Millimeter-Wave Dielectric Haloscope}},
  journal={Phys. Rev. Lett.}, volume={136}, pages={071001}, year={2026}, doi={10.1103/yk7n-lrpj}, eprint={2503.23354}, archivePrefix={arXiv}}

@misc{Ourahou:2026dandelion,
    author = "Ourahou, I. and Savorgnano, S. and Beaufort, C. and Bastero-Gil, M. and Bounmy, J. and others",
    title = "{First Dark Photon Search Results from the Dandelion Experiment}",
    eprint = "2602.18218",
    archivePrefix = "arXiv",
    primaryClass = "astro-ph.CO",
    month = "2",
    year = "2026"
}

@misc{ALPHA:2026phase1,
    author = "Bai, Xiran and Balafendiev, Rustam and others",
    collaboration = "ALPHA",
    title = "{Design of ALPHA Phase I: A Plasma Haloscope for 10--20 GHz Post-Inflation Axions}",
    eprint = "2608.10292",
    archivePrefix = "arXiv",
    primaryClass = "hep-ex",
    month = "8",
    year = "2026"
}

@misc{SM,
  note={See Supplemental Material, appended below, for the circular-wire and full-wave validation, the finite-tile collection, the mechanical tolerances, the halo dephasing factor, and the overlap formulation of the boost, which includes Refs.~\cite{Brillouin:1946,Collin:1991,Nicorovici:1995,Botten:2000,Linton:1998,Yee:1966,WiregridGitHub,Ioannisian:2017srr,Egge:2022gfp,Egge:2023cos,Garutti:2026asa}.}}

@article{Mok:1979,
  author = {Mok, C. L. and Chambers, W. G. and Parker, T. J. and Costley, A. E.},
  title = {{The far-infrared performance and application of free-standing grids wound from 5 $\mu$m diameter tungsten wire}},
  journal = {Infrared Phys.},
  volume = {19}, pages = {437}, year = {1979},
  doi = {10.1016/0020-0891(79)90055-1}
}

@article{Ioannisian:2017srr,
    author = "Ioannisian, Ara N. and Kazarian, Narine and Millar, Alexander J. and Raffelt, Georg G.",
    title = "{Axion-photon conversion caused by dielectric interfaces: quantum field calculation}",
    eprint = "1707.00701",
    archivePrefix = "arXiv",
    primaryClass = "hep-ph",
    doi = "10.1088/1475-7516/2017/09/005",
    journal = "JCAP",
    volume = "2017",
    number = "09",
    pages = "005",
    year = "2017"
}

@article{ADMX:2018ogs,
    author = "Boutan, C. and others",
    collaboration = "ADMX",
    title = "{Piezoelectrically Tuned Multimode Cavity Search for Axion Dark Matter}",
    eprint = "1901.00920",
    archivePrefix = "arXiv",
    primaryClass = "hep-ex",
    reportNumber = "FERMILAB-PUB-18-702-AD-AE",
    doi = "10.1103/PhysRevLett.121.261302",
    journal = "Phys. Rev. Lett.",
    volume = "121",
    number = "26",
    pages = "261302",
    year = "2018"
}

@article{ADMX:2021mio,
    author = "Bartram, C. and others",
    collaboration = "ADMX",
    title = "{Dark matter axion search using a Josephson Traveling wave parametric amplifier}",
    eprint = "2110.10262",
    archivePrefix = "arXiv",
    primaryClass = "hep-ex",
    reportNumber = "FERMILAB-PUB-22-615-PPD-SQMS",
    doi = "10.1063/5.0122907",
    journal = "Rev. Sci. Instrum.",
    volume = "94",
    number = "4",
    pages = "044703",
    year = "2023"
}

@article{Jeong:2020cwz,
    author = "Jeong, Junu and Youn, SungWoo and Bae, Sungjae and Kim, Jihngeun and Seong, Taehyeon and Kim, Jihn E. and Semertzidis, Yannis K.",
    title = "{Search for Invisible Axion Dark Matter with a Multiple-Cell Haloscope}",
    eprint = "2008.10141",
    archivePrefix = "arXiv",
    primaryClass = "hep-ex",
    doi = "10.1103/PhysRevLett.125.221302",
    journal = "Phys. Rev. Lett.",
    volume = "125",
    number = "22",
    pages = "221302",
    year = "2020"
}

@article{Lee:2022mnc,
    author = "Lee, Youngjae and Yang, Byeongsu and Yoon, Hojin and Ahn, Moohyun and Park, Heejun and Min, Byeonghun and Kim, DongLak and Yoo, Jonghee",
    title = "{Searching for Invisible Axion Dark Matter with an 18~T Magnet Haloscope}",
    eprint = "2206.08845",
    archivePrefix = "arXiv",
    primaryClass = "hep-ex",
    doi = "10.1103/PhysRevLett.128.241805",
    journal = "Phys. Rev. Lett.",
    volume = "128",
    number = "24",
    pages = "241805",
    year = "2022"
}

@article{Yang:2023yry,
    author = "Yang, Byeongsu and Yoon, Hojin and Ahn, Moohyun and Lee, Youngjae and Yoo, Jonghee",
    title = "{Extended Axion Dark Matter Search Using the CAPP18T Haloscope}",
    eprint = "2308.09077",
    archivePrefix = "arXiv",
    primaryClass = "hep-ex",
    doi = "10.1103/PhysRevLett.131.081801",
    journal = "Phys. Rev. Lett.",
    volume = "131",
    number = "8",
    pages = "081801",
    year = "2023"
}

@article{Kim:2023vpo,
    author = "Kim, Younggeun and others",
    title = "{Experimental Search for Invisible Dark Matter Axions around 22~$\mu$eV}",
    eprint = "2312.11003",
    archivePrefix = "arXiv",
    primaryClass = "hep-ex",
    doi = "10.1103/PhysRevLett.133.051802",
    journal = "Phys. Rev. Lett.",
    volume = "133",
    number = "5",
    pages = "051802",
    year = "2024"
}

@article{Bae:2024kmy,
    author = "Bae, Sungjae and Jeong, Junu and Kim, Younggeun and Youn, SungWoo and Park, Heejun and Seong, Taehyeon and Oh, Seongjeong and Semertzidis, Yannis K.",
    title = "{Search for Dark Matter Axions with Tunable TM020 Mode}",
    eprint = "2403.13390",
    archivePrefix = "arXiv",
    primaryClass = "hep-ex",
    doi = "10.1103/PhysRevLett.133.211803",
    journal = "Phys. Rev. Lett.",
    volume = "133",
    number = "21",
    pages = "211803",
    year = "2024"
}

@article{CAPP:2020utb,
    author = "Kwon, Ohjoon and others",
    collaboration = "CAPP",
    title = "{First Results from an Axion Haloscope at CAPP around 10.7~$\mu$eV}",
    eprint = "2012.10764",
    archivePrefix = "arXiv",
    primaryClass = "hep-ex",
    doi = "10.1103/PhysRevLett.126.191802",
    journal = "Phys. Rev. Lett.",
    volume = "126",
    number = "19",
    pages = "191802",
    year = "2021"
}

@article{Adair:2022rtw,
    author = "Adair, C. M. and others",
    title = "{Search for Dark Matter Axions with CAST-CAPP}",
    eprint = "2211.02902",
    archivePrefix = "arXiv",
    primaryClass = "hep-ex",
    doi = "10.1038/s41467-022-33913-6",
    journal = "Nature Commun.",
    volume = "13",
    number = "1",
    pages = "6180",
    year = "2022"
}

@misc{Grenet:2021vbb,
    author = "Grenet, Thierry and Ballou, Rafik and Basto, Quentin and Martineau, Killian and Perrier, Pierre and Pugnat, Pierre and Quevillon, J{\'e}r{\'e}mie and Roch, Nicolas and Smith, Christopher",
    title = "{The Grenoble Axion Haloscope platform (GrAHal): development plan and first results}",
    eprint = "2110.14406",
    archivePrefix = "arXiv",
    primaryClass = "hep-ex",
    month = "10",
    year = "2021"
}

@article{Brubaker:2016ktl,
    author = "Brubaker, B. M. and others",
    title = "{First results from a microwave cavity axion search at 24 $\mu$eV}",
    eprint = "1610.02580",
    archivePrefix = "arXiv",
    primaryClass = "astro-ph.CO",
    doi = "10.1103/PhysRevLett.118.061302",
    journal = "Phys. Rev. Lett.",
    volume = "118",
    number = "6",
    pages = "061302",
    year = "2017"
}

@article{HAYSTAC:2018rwy,
    author = "Zhong, L. and others",
    collaboration = "HAYSTAC",
    title = "{Results from phase 1 of the HAYSTAC microwave cavity axion experiment}",
    eprint = "1803.03690",
    archivePrefix = "arXiv",
    primaryClass = "hep-ex",
    doi = "10.1103/PhysRevD.97.092001",
    journal = "Phys. Rev. D",
    volume = "97",
    number = "9",
    pages = "092001",
    year = "2018"
}

@article{HAYSTAC:2023cam,
    author = "Jewell, M. J. and others",
    collaboration = "HAYSTAC",
    title = "{New results from HAYSTAC{\textquoteright}s phase II operation with a squeezed state receiver}",
    eprint = "2301.09721",
    archivePrefix = "arXiv",
    primaryClass = "hep-ex",
    doi = "10.1103/PhysRevD.107.072007",
    journal = "Phys. Rev. D",
    volume = "107",
    number = "7",
    pages = "072007",
    year = "2023"
}

@article{HAYSTAC:2024jch,
    author = "Bai, Xiran and others",
    collaboration = "HAYSTAC",
    title = "{Dark Matter Axion Search with HAYSTAC Phase II}",
    eprint = "2409.08998",
    archivePrefix = "arXiv",
    primaryClass = "hep-ex",
    doi = "10.1103/PhysRevLett.134.151006",
    journal = "Phys. Rev. Lett.",
    volume = "134",
    number = "15",
    pages = "151006",
    year = "2025"
}

@article{McAllister:2017lkb,
    author = "McAllister, Ben T. and Flower, Graeme and Kruger, Justin and Ivanov, Eugene N. and Goryachev, Maxim and Bourhill, Jeremy and Tobar, Michael E.",
    collaboration = "ORGAN",
    title = "{The ORGAN Experiment: An axion haloscope above 15 GHz}",
    eprint = "1706.00209",
    archivePrefix = "arXiv",
    primaryClass = "physics.ins-det",
    doi = "10.1016/j.dark.2017.09.010",
    journal = "Phys. Dark Univ.",
    volume = "18",
    pages = "67--72",
    year = "2017"
}

@article{Alesini:2019ajt,
    author = "Alesini, D. and others",
    title = "{Galactic axions search with a superconducting resonant cavity}",
    eprint = "1903.06547",
    archivePrefix = "arXiv",
    primaryClass = "physics.ins-det",
    doi = "10.1103/PhysRevD.99.101101",
    journal = "Phys. Rev. D",
    volume = "99",
    number = "10",
    pages = "101101",
    year = "2019"
}

@article{Alesini:2020vny,
    author = "Alesini, D. and others",
    title = "{Search for invisible axion dark matter of mass $m_a=43~\mu$eV with the QUAX--$a\gamma$ experiment}",
    eprint = "2012.09498",
    archivePrefix = "arXiv",
    primaryClass = "hep-ex",
    doi = "10.1103/PhysRevD.103.102004",
    journal = "Phys. Rev. D",
    volume = "103",
    number = "10",
    pages = "102004",
    year = "2021"
}

@article{Alesini:2022lnp,
    author = "Alesini, D. and others",
    title = "{Search for Galactic axions with a high-Q dielectric cavity}",
    eprint = "2208.12670",
    archivePrefix = "arXiv",
    primaryClass = "hep-ex",
    doi = "10.1103/PhysRevD.106.052007",
    journal = "Phys. Rev. D",
    volume = "106",
    number = "5",
    pages = "052007",
    year = "2022"
}

@article{QUAX:2023gop,
    author = "Di Vora, R. and others",
    collaboration = "QUAX",
    title = "{Search for galactic axions with a traveling wave parametric amplifier}",
    eprint = "2304.07505",
    archivePrefix = "arXiv",
    primaryClass = "hep-ex",
    reportNumber = "FERMILAB-PUB-23-229-SQMS-V",
    doi = "10.1103/PhysRevD.108.062005",
    journal = "Phys. Rev. D",
    volume = "108",
    number = "6",
    pages = "062005",
    year = "2023"
}

@article{CAST:2020rlf,
    author = "{\'A}lvarez Melc{\'o}n, A. and others",
    collaboration = "CAST",
    title = "{First results of the CAST-RADES haloscope search for axions at 34.67 $\mu$eV}",
    eprint = "2104.13798",
    archivePrefix = "arXiv",
    primaryClass = "hep-ex",
    reportNumber = "CERN-EP-2021-070",
    doi = "10.1007/JHEP10(2021)075",
    journal = "JHEP",
    volume = "2021",
    number = "10",
    pages = "075",
    year = "2021"
}

@article{Ahyoune:2024klt,
    author = "Ahyoune, S. and others",
    title = "{RADES axion search results with a high-temperature superconducting cavity in an 11.7 T magnet}",
    eprint = "2403.07790",
    archivePrefix = "arXiv",
    primaryClass = "hep-ex",
    reportNumber = "CERN-EP-2024-076, MPP-2024-55",
    doi = "10.1007/JHEP04(2025)113",
    journal = "JHEP",
    volume = "2025",
    number = "04",
    pages = "113",
    year = "2025"
}

@article{TASEH:2022vvu,
    author = "Chang, Hsin and others",
    collaboration = "TASEH",
    title = "{First Results from the Taiwan Axion Search Experiment with a Haloscope at 19.6~$\mu$eV}",
    eprint = "2205.05574",
    archivePrefix = "arXiv",
    primaryClass = "hep-ex",
    doi = "10.1103/PhysRevLett.129.111802",
    journal = "Phys. Rev. Lett.",
    volume = "129",
    number = "11",
    pages = "111802",
    year = "2022"
}

@article{Wuensch:1989sa,
    author = "Wuensch, Walter and De Panfilis-Wuensch, S. and Semertzidis, Y. K. and Rogers, J. T. and Melissinos, A. C. and Halama, H. J. and Moskowitz, B. E. and Prodell, A. G. and Fowler, W. B. and Nezrick, F. A.",
    title = "{Results of a Laboratory Search for Cosmic Axions and Other Weakly Coupled Light Particles}",
    reportNumber = "FERMILAB-PUB-89-185-E, BNL-43010",
    doi = "10.1103/PhysRevD.40.3153",
    journal = "Phys. Rev. D",
    volume = "40",
    pages = "3153",
    year = "1989"
}

@article{An:2020bxd,
    author = "An, Haipeng and Pospelov, Maxim and Pradler, Josef and Ritz, Adam",
    title = "{New limits on dark photons from solar emission and keV scale dark matter}",
    eprint = "2006.13929",
    archivePrefix = "arXiv",
    primaryClass = "hep-ph",
    doi = "10.1103/PhysRevD.102.115022",
    journal = "Phys. Rev. D",
    volume = "102",
    pages = "115022",
    year = "2020"
}

@article{Witte:2020rvb,
    author = "Witte, Samuel J. and Rosauro-Alcaraz, Salvador and McDermott, Samuel D. and Poulin, Vivian",
    title = "{Dark photon dark matter in the presence of inhomogeneous structure}",
    eprint = "2003.13698",
    archivePrefix = "arXiv",
    primaryClass = "astro-ph.CO",
    reportNumber = "FERMILAB-PUB-20-121-T",
    doi = "10.1007/JHEP06(2020)132",
    journal = "JHEP",
    volume = "2020",
    number = "06",
    pages = "132",
    year = "2020"
}

@article{Caputo:2020bdy,
    author = "Caputo, Andrea and Liu, Hongwan and Mishra-Sharma, Siddharth and Ruderman, Joshua T.",
    title = "{Dark Photon Oscillations in Our Inhomogeneous Universe}",
    eprint = "2002.05165",
    archivePrefix = "arXiv",
    primaryClass = "astro-ph.CO",
    doi = "10.1103/PhysRevLett.125.221303",
    journal = "Phys. Rev. Lett.",
    volume = "125",
    number = "22",
    pages = "221303",
    year = "2020"
}

@article{Bajjali:2023uis,
    author = "Bajjali, Fayez and others",
    title = "{First results from BRASS-p broadband searches for hidden photon dark matter}",
    eprint = "2306.05934",
    archivePrefix = "arXiv",
    primaryClass = "hep-ex",
    doi = "10.1088/1475-7516/2023/08/077",
    journal = "JCAP",
    volume = "2023",
    number = "08",
    pages = "077",
    year = "2023"
}

@article{DOSUE-RR:2022ise,
    author = "Kotaka, Shumpei and others",
    collaboration = "DOSUE-RR",
    title = "{Search for Dark Photon Dark Matter in the Mass Range 74--110~$\mu$eV with a Cryogenic Millimeter-Wave Receiver}",
    eprint = "2205.03679",
    archivePrefix = "arXiv",
    primaryClass = "hep-ex",
    doi = "10.1103/PhysRevLett.130.071805",
    journal = "Phys. Rev. Lett.",
    volume = "130",
    number = "7",
    pages = "071805",
    year = "2023"
}

@article{DOSUE-RR:2023yml,
    author = "Adachi, S. and Fujinaka, R. and Muto, Y. and Nakata, H. and Sueno, Y. and Sumida, T. and Suzuki, J. and Tajima, O. and Takeuchi, H. and Honda, S.",
    collaboration = "DOSUE-RR",
    title = "{Search for dark photon dark matter in the mass range 41--74~$\mu$eV using millimeter-wave receiver and radioshielding box}",
    eprint = "2308.14656",
    archivePrefix = "arXiv",
    primaryClass = "hep-ex",
    doi = "10.1103/PhysRevD.109.012008",
    journal = "Phys. Rev. D",
    volume = "109",
    number = "1",
    pages = "012008",
    year = "2024"
}

@article{MADMAX:2024jnp,
    author = "Egge, J. and others",
    collaboration = "MADMAX",
    title = "{First Search for Dark Photon Dark Matter with a madmax Prototype}",
    eprint = "2408.02368",
    archivePrefix = "arXiv",
    primaryClass = "hep-ex",
    reportNumber = "FERMILAB-PUB-24-0466-PPD",
    doi = "10.1103/PhysRevLett.134.151004",
    journal = "Phys. Rev. Lett.",
    volume = "134",
    number = "15",
    pages = "151004",
    year = "2025"
}

@article{Cervantes:2022yzp,
    author = "Cervantes, R. and others",
    title = "{Search for 70~$\mu$eV Dark Photon Dark Matter with a Dielectrically Loaded Multiwavelength Microwave Cavity}",
    eprint = "2204.03818",
    archivePrefix = "arXiv",
    primaryClass = "hep-ex",
    reportNumber = "FERMILAB-PUB-22-264-V",
    doi = "10.1103/PhysRevLett.129.201301",
    journal = "Phys. Rev. Lett.",
    volume = "129",
    number = "20",
    pages = "201301",
    year = "2022"
}

@article{Ramanathan:2022egk,
    author = "Ramanathan, Karthik and Klimovich, Nikita and Basu Thakur, Ritoban and Eom, Byeong Ho and LeDuc, Henry G. and Shu, Shibo and Beyer, Andrew D. and Day, Peter K.",
    title = "{Wideband Direct Detection Constraints on Hidden Photon Dark Matter with the QUALIPHIDE Experiment}",
    eprint = "2209.03419",
    archivePrefix = "arXiv",
    primaryClass = "astro-ph.CO",
    doi = "10.1103/PhysRevLett.130.231001",
    journal = "Phys. Rev. Lett.",
    volume = "130",
    number = "23",
    pages = "231001",
    year = "2023"
}

@article{Brun:2019kak,
    author = "Brun, Pierre and Chevalier, Laurent and Flouzat, Christophe",
    title = "{Direct Searches for Hidden-Photon Dark Matter with the SHUKET Experiment}",
    eprint = "1905.05579",
    archivePrefix = "arXiv",
    primaryClass = "hep-ex",
    doi = "10.1103/PhysRevLett.122.201801",
    journal = "Phys. Rev. Lett.",
    volume = "122",
    number = "20",
    pages = "201801",
    year = "2019"
}

@misc{Schneemann:2023bqc,
    author = "Schneemann, Tim and Schmieden, Kristof and Schott, Matthias",
    title = "{First results of the SUPAX Experiment: Probing Dark Photons}",
    eprint = "2308.08337",
    archivePrefix = "arXiv",
    primaryClass = "hep-ex",
    month = "8",
    year = "2023"
}

@article{Tomita:2020usq,
    author = "Tomita, Nozomu and Oguri, Shugo and Inoue, Yoshizumi and Minowa, Makoto and Nagasaki, Taketo and Suzuki, Jun'ya and Tajima, Osamu",
    title = "{Search for hidden-photon cold dark matter using a K-band cryogenic receiver}",
    eprint = "2006.02828",
    archivePrefix = "arXiv",
    primaryClass = "hep-ex",
    doi = "10.1088/1475-7516/2020/09/012",
    journal = "JCAP",
    volume = "2020",
    number = "09",
    pages = "012",
    year = "2020"
}

@book{Collin:1991,
    author    = "Collin, Robert E.",
    title     = "{Field Theory of Guided Waves}",
    edition   = "2nd",
    publisher = "IEEE Press",
    address   = "New York",
    year      = "1991"
}

@book{Brillouin:1946,
    author    = "Brillouin, L{\'e}on",
    title     = "{Wave Propagation in Periodic Structures}",
    publisher = "McGraw-Hill",
    address   = "New York",
    year      = "1946"
}

@article{Botten:2000,
    author  = "Botten, L. C. and Nicorovici, N. A. P. and Asatryan, A. A. and McPhedran, R. C. and de Sterke, C. M. and Robinson, P. A.",
    title   = "{Formulation for electromagnetic scattering and propagation through grating stacks of metallic and dielectric cylinders for photonic crystal calculations. Part I. Method}",
    journal = "J. Opt. Soc. Am. A",
    volume  = "17",
    pages   = "2165",
    year    = "2000",
    doi     = "10.1364/JOSAA.17.002165"
}

@article{Nicorovici:1995,
    author  = "Nicorovici, N. A. and McPhedran, R. C. and Botten, L. C.",
    title   = "{Photonic band gaps for arrays of perfectly conducting cylinders}",
    journal = "Phys. Rev. E",
    volume  = "52",
    pages   = "1135--1145",
    year    = "1995",
    doi     = "10.1103/PhysRevE.52.1135"
}

@article{Linton:1998,
    author  = "Linton, C. M.",
    title   = "{The Green's function for the two-dimensional Helmholtz equation in periodic domains}",
    journal = "J. Eng. Math.",
    volume  = "33",
    pages   = "377--401",
    year    = "1998",
    doi     = "10.1023/A:1004377501747"
}

@misc{WiregridGitHub,
  author = {Caputo, Andrea},
  title = {{Wiregrid\_Haloscope: codes and data for the layered wire-grid haloscope}},
  howpublished = {\url{https://github.com/andrea0292/Wiregrid_Haloscope}},
  year = {2026}
}

@article{Yee:1966,
    author = "Yee, Kane S.",
    title = "{Numerical solution of initial boundary value problems involving Maxwell's equations in isotropic media}",
    journal = "IEEE Trans. Antennas Propag.",
    volume = "14",
    number = "3",
    pages = "302--307",
    year = "1966",
    doi = "10.1109/TAP.1966.1138693"
}

@article{Benabou:2024msj,
    author = "Benabou, Joshua N. and Buschmann, Malte and Foster, Joshua W. and Safdi, Benjamin R.",
    title = "{Axion Mass Prediction from Adaptive Mesh Refinement Cosmological Lattice Simulations}",
    eprint = "2412.08699",
    archivePrefix = "arXiv",
    primaryClass = "hep-ph",
    reportNumber = "FERMILAB-PUB-24-0912-T",
    doi = "10.1103/6v21-d6sj",
    journal = "Phys. Rev. Lett.",
    volume = "134",
    number = "24",
    pages = "241003",
    year = "2025"
}

@misc{Garutti:2026asa,
    author = "Garutti, Erika and Ivanov, Anton and Majorovits, B{\'e}la",
    title = "{Comment on the preprint: First Limits on Axion Dark Matter from a DALI Prototype arXiv:2603.21951}",
    eprint = "2607.28095",
    archivePrefix = "arXiv",
    primaryClass = "hep-ex",
    year = "2026"
}

@article{Safdi:2022xkm,
    author = "Safdi, Benjamin R.",
    title = "{TASI Lectures on the Particle Physics and Astrophysics of Dark Matter}",
    eprint = "2303.02169",
    archivePrefix = "arXiv",
    primaryClass = "hep-ph",
    doi = "10.22323/1.439.0009",
    journal = "PoS",
    volume = "TASI2022",
    pages = "009",
    year = "2024"
}

@article{OHare:2024nmr,
    author = "O'Hare, Ciaran A. J.",
    title = "{Cosmology of axion dark matter}",
    eprint = "2403.17697",
    archivePrefix = "arXiv",
    primaryClass = "hep-ph",
    doi = "10.22323/1.454.0040",
    journal = "PoS",
    volume = "COSMICWISPers",
    pages = "040",
    year = "2024"
}

@article{Caputo:2024oqc,
    author = "Caputo, Andrea and Raffelt, Georg",
    title = "{Astrophysical Axion Bounds: The 2024 Edition}",
    eprint = "2401.13728",
    archivePrefix = "arXiv",
    primaryClass = "hep-ph",
    reportNumber = "MPP-2024-13, CERN-TH-2024-013",
    doi = "10.22323/1.454.0041",
    journal = "PoS",
    volume = "COSMICWISPers",
    pages = "041",
    year = "2024"
}

\clearpage
\onecolumngrid
\begin{center}{\large\bfseries End Matter}\end{center}
\twocolumngrid
\setcounter{equation}{0}
\renewcommand{\theequation}{A\arabic{equation}}
\renewcommand{\theHequation}{A.\arabic{equation}}

\section*{A. Resonant emission from the wire-grid slab}

In this section we derive the slab response used in the main text. Throughout, $\varepsilon$ denotes the permittivity for fields along the wires, which are parallel to $\mathbf B_0$. When the grid separation $d$ is small compared with the wavelength, the stack can be described by averaging the grid currents over that separation. A grid carrying the surface current $K=E_\parallel/Z_g$, where $E_\parallel$ is the field along the wires, then corresponds to a volume current density $J=E_\parallel/(dZ_g)$, so the stack behaves as a medium of conductivity $1/(dZ_g)$. Inserting $Z_g=R_g-i\omega L_g$ gives the Drude permittivity of Eq.~\eqref{eq:plasma}, with $\omega_p^2=1/(dL_g)$ and $\Gamma=R_g/L_g$. For a spatially uniform axion source with time dependence $e^{-i\omega t}$, Amp\`ere's law becomes $-i\omega\varepsilon\mathbf E+\mathbf J_a=0$, where $\mathbf J_a=\gagg\dot a\,\mathbf B_0$, and the particular solution is then $\mathbf E=-\gagg a\mathbf B_0/\varepsilon$, which explains the enhancement near $\varepsilon=0$. However, a uniform particular solution alone carries no outgoing radiation. The emitted wave must be obtained by imposing the boundary conditions on the driven field and the electromagnetic waves that accompany it.

Let $s$ measure distance normal to a slab occupying $0<s<L$, with a perfect mirror at $s=0$. For the source confined to the slab, the fields along the wires can be written as
\begin{equation}
\begin{aligned}
E_{\rm in}(s)&=\frac{E_0}{\varepsilon}\big[\cos(nks)-1\big]
                +C\sin(nks),\\
E_{\rm out}(s)&=\beta E_0e^{ik(s-L)},
\end{aligned}
\end{equation}
where $n=\sqrt{\varepsilon}$, $k=\omega$, $E_0=\gagg a_0B_0$, and $C$ is a wave amplitude fixed by matching at the face. This form already enforces $E_{\rm in}(0)=0$. Since the permeability is the same on both sides, continuity of the tangential electric and magnetic fields requires $E_{\rm in}(L)=\beta E_0$ and $E'_{\rm in}(L)=ik\beta E_0$, where a prime denotes $\dd/\dd s$. Eliminating $C$ gives
\begin{equation}
\beta=\frac{1-\cos\phi}{\varepsilon\cos\phi-in\sin\phi},
\qquad \phi=nkL.
\label{eq:betaslab}
\end{equation}
Thus $|\beta|^2$ is the emitted power relative to a bare mirror of the same area. For a lossless medium with $\varepsilon=u>0$, the half-wave condition $\phi=\pi$ gives $L=\lambda/(2\sqrt{u})$ and $\beta=-2/u$, where $\lambda=2\pi/k$ is the vacuum wavelength.

Absorption regularizes this enhancement. Slightly above the plasma frequency, write $\varepsilon\simeq u+i/\Qm$, with $\Qm=\omega_p/\Gamma$ and $\Qm^{-1}\ll u\ll1$. At the lossless half-wave condition, the phase becomes $\phi\simeq\pi+i\pi/(2u\Qm)$, so the dominant real part of the denominator in Eq.~\eqref{eq:betaslab} is $-[u+\pi/(2\Qm\sqrt{u})]$. Writing $x=\gamma_{\rm rad}/\gamma_{\rm ohm}$ for the ratio of radiative to ohmic energy-decay rates, the peak amplitude is
\begin{equation}
|\beta_{\rm pk}|\simeq\frac{2}{u+\pi/(2\Qm\sqrt{u})}
=\frac{2}{u}\frac{x}{1+x},
\label{eq:peakescape}
\end{equation}
and squaring this expression gives Eq.~\eqref{eq:betamax}.

To identify these rates, let $U$ be the stored energy, which decays as $\dot U=-(\gamma_{\rm rad}+\gamma_{\rm ohm})U$ after the drive is removed. For the lossless Drude law, $nk=\sqrt{\omega^2-\omega_p^2}$, so the group velocity $v_g=[\dd(nk)/\dd\omega]^{-1}=\sqrt{u}$ becomes small near the plasma frequency. The power transmission through the face is then $\mathcal T=4\sqrt{u}/(1+\sqrt{u})^2\simeq4\sqrt{u}$. A wave encounters that face once per round trip of duration $2L/v_g$, giving $\gamma_{\rm rad}\simeq\mathcal T v_g/(2L)=2\omega u^{3/2}/\pi$. Together with $\gamma_{\rm ohm}\simeq\Gamma$, this yields
\begin{equation}
x\equiv\frac{\gamma_{\rm rad}}{\gamma_{\rm ohm}}
\simeq\frac{2\Qm u^{3/2}}{\pi},
\qquad \Qr\simeq\frac{\Qm}{1+x}.
\end{equation}
Here $\Qr=\nu_{\rm res}/\Delta\nu_{\rm FWHM}$ describes the full width at half maximum of the power response; the corresponding angular-frequency width is $\gamma_{\rm rad}+\gamma_{\rm ohm}$. Equation~\eqref{eq:peakescape} therefore expresses the lossless enhancement multiplied by the fraction of stored energy that escapes. For the discrete stacks of Table~\ref{tab:design}, whose gaps are not small compared with the wavelength, the ohmic quality factor $Q_{\rm ohm}\equiv\omega/\gamma_{\rm ohm}$ differs from the material parameter $\Qm$. The calculated linewidths give $Q_{\rm ohm}\simeq(1.2$--$1.3)\,\Qm$, with $\Qr\simeq Q_{\rm ohm}/(1+x)$.

A scan also benefits from the bandwidth of the enhancement. In the background-dominated regime the squared signal-to-noise ratio grows as the integration time times the squared signal power, and hence as $|\beta|^4$. Writing $\nu_{\rm res}$ for the tuned resonance frequency, uniform exposure per logarithmic interval means $\dd t=(T_{\rm oct}/\ln2)\,\dd\ln\nu_{\rm res}$, where $T_{\rm oct}$ is the live time per octave. Combining measurements as the resonance passes a fixed axion frequency gives a squared signal-to-noise ratio proportional to $\mathcal I=\int|\beta(\nu)|^4\,\dd\nu/\nu$, provided the detector noise and response shape vary slowly across the resonance. For a narrow Lorentzian,
\begin{equation}
\begin{aligned}
|\beta(\nu)|^2&\simeq\frac{|\beta_{\rm pk}|^2}
 {1+4\Qr^2[(\nu-\nu_{\rm res})/\nu_{\rm res}]^2},\\
\mathcal I&\simeq\frac{\pi|\beta_{\rm pk}|^4}{4\Qr}
\propto\frac{x^{4/3}}{(1+x)^3}.
\end{aligned}
\end{equation}
This expression is maximal at $x=4/5$, and it is this scan optimum that guides the choice of the number of grids. In the forecasts for power detectors we conservatively count only the exposure during which the axion frequency lies within the half-maximum width of the response. For a Lorentzian this retains a fraction $1/2+1/\pi\simeq0.82$ of $\mathcal I$, and it raises the coupling threshold by about 5\%.

For the actual stack, the vacuum gaps are not small compared with the wavelength, so the slab result should be seen only as a scaling guide. In each gap we retain the driven field and both traveling waves, impose continuity of the electric field and the magnetic-field jump due to the sheet current, and apply the mirror and outgoing-wave conditions. Solving this linear system gives $\beta(\nu)$ and $\mathcal I$ without homogenization. Radiative and ohmic linewidths are separated by repeating the calculation with negligible resistance, which gives the radiative quality factor $Q_{\rm rad}$ and $Q_{\rm ohm}^{-1}=\Qr^{-1}-Q_{\rm rad}^{-1}$; the circular-wire checks and the correction for a source extending outside the stack are detailed in~\cite{SM}.

\renewcommand{\theequation}{B\arabic{equation}}
\renewcommand{\theHequation}{B.\arabic{equation}}
\setcounter{equation}{0}
\section*{B. Comparison with other haloscopes}

MADGRID targets 0.25--8\,meV, above the masses addressed by wire-plasma haloscopes such as ALPHA~\cite{Lawson:2019brd,ALPHA:2022rxj}, whose first phase targets 10--20\,GHz~\cite{ALPHA:2026phase1}, and by the dielectric haloscopes MADMAX and DALI~\cite{MADMAX:2019pub,DeMiguel:2020rpn}, which aim at masses up to a few hundred $\mu$eV. Geometrically, the closest of these are MADMAX and DALI, which are also mirror-backed layered boosters, and although the area law for lossless dielectric disks~\cite{Millar:2016cjp} does not directly apply to dispersive and lossy grids, integrating over 0.5--1.5 times the design frequency gives $\int|\beta|^2\dd\nu/\nu\simeq2.5$ per sapphire disk ($\varepsilon=9.4$, half-wave configuration) and $\simeq2.6$ per grid for our lossless 1--2\,meV design, or $\simeq1.3$ with copper losses, so the two boosters are comparable per element. They differ in how they sculpt the mode, because disks half a wavelength apart have a refractive index above unity, whereas grids a quarter wavelength apart act as inductive sheets whose stack, just above its band edge, has an effective index close to zero, the ENZ regime that SQWARE reaches at similar masses with quantum wells in much stronger fields~\cite{Mehrani:2025ryc}.

Comparing the reach of different concepts at a given mass requires the signal power delivered to each detector, its noise, and the observing time spent at that mass. For independent power detectors with white noise, the squared signal-to-noise ratio is
\begin{equation}
\mathrm{SNR}^2=2\sum_{i,j}
\frac{t_i P_{s,ij}^{\,2}}{{\rm NEP}_j^{\,2}},
\label{eq:comparison_snr}
\end{equation}
where $t_i$ is the live time at tuning setting $i$ and $P_{s,ij}$ the signal power detected by detector $j$, and the same expression holds for background-dominated photon counting upon replacing ${\rm NEP}_j^2$ by $2(2\pi\nu)^2D_j$, with $D_j$ the background count rate. Dividing a fixed total signal equally among $N_{\rm det}$ independent detectors therefore reduces $\mathrm{SNR}^2$ by $N_{\rm det}$ relative to collecting it onto a single detector with the same noise and efficiency, whenever such collection is possible.

For axions at fixed mass and density, the signal power scales as $\gagg^2 B_0^2 A\eta|\beta|^2$, where $\eta=\eta_d\eta_{\rm ap}\xi$ includes collection, transmission and detection efficiency. A scan must also account for the resonance bandwidth, as a larger peak boost need not improve sensitivity if each mass is observed for a correspondingly shorter time. Uniform exposure per logarithmic frequency interval introduces the response integral $\mathcal I$ derived in End Matter A. For identical channels sharing total area $A$, at fixed noise per detector, the coupling reach therefore scales as
\begin{equation}
g_{\min}\propto
\left[
\frac{(\eta B_0^2A)^2\,T_{\rm oct}\mathcal I}
{N_{\rm det}}
\right]^{-1/4}.
\end{equation}

SQWARE uses the same radiative idea with a smaller area, a stronger resonance and a larger field. For its LM and LM-S configurations~\cite{Mehrani:2025ryc}, resonant at 1.53 and 1.07\,meV, we find $\mathcal I\simeq8.5\times10^3$ and $1.0\times10^5$ with $B_0^2A=1.17$ and $6.25\,{\rm T^2m^2}$, against our 100 tiles at $B_0^2A=100\,{\rm T^2m^2}$. Assuming unit collection onto one detector per SQWARE sample, and equal detector noise and time per logarithmic interval, our array reaches couplings 2.4 times smaller than LM, while LM-S reaches couplings 1.9 times smaller than ours. SQWARE, however, requires 2--20\,mm quantum-well structures and 36--50\,T fields, whereas the grids require validation of copper losses, tolerances and collection, and operate without a magnet for DP. For ALPHA the geometry and the extraction of the signal both differ~\cite{Lawson:2019brd,ALPHA:2022rxj}. ALPHA fills the bore of a solenoid with a three-dimensional wire metamaterial enclosed in a resonator, tuned by moving the wires of the lattice and read out through an antenna coupled to the resonator, with its first phase targeting 10--20\,GHz~\cite{ALPHA:2026phase1}. At higher frequencies, ohmic attenuation limits the region that one localized antenna can collect efficiently. Additional antennas can recover signal from a larger volume, as discussed by ALPHA~\cite{ALPHA:2022rxj}, but their signals must then be transported to receivers or combined via a collection network. 

Finally, for DPs BREAD projects $\chi\simeq3\times10^{-14}$ for a $10\,{\rm m^2}$ dish read by one detector with ${\rm NEP}=2\times10^{-19}\,{\rm W/\sqrt{Hz}}$ for 1000\,d~\cite{BREAD:2021tpx}. Our $1\,{\rm m^2}$ KID forecast is about four times weaker near 1\,meV, because it uses a tenth of the area, 100 independent detectors, each adding its own noise, and a three times larger NEP, and the resonance recovers only part of this loss. To compare the two on equal footing we use Eq.~\eqref{eq:comparison_snr}, noting that a dish observes every mass with $|\beta|=1$ for its full live time $t$, whereas our scan spends $t/5$ on each octave and weights each mass by the scan integral $\mathcal I$ of End Matter~A. At equal area, detector noise and number of detectors, the ratio of squared signal-to-noise ratios is then $(\mathcal I/5\ln2)(\kappa_{\rm w}/\kappa_{\rm d})^2(\eta_{\rm ap}\xi)^2$, with $\kappa_{\rm w}/\kappa_{\rm d}=1/2$ because the wires accept only the polarization along them, whereas a dish responds to both components in its surface. Since $\mathcal I\simeq\pi|\beta_{\rm pk}|^4/(4\Qr)$ ranges from $2.6\times10^3$ to $4\times10^4$ for the cartridges of Table~\ref{tab:design}, and the reach in $\chi$ scales as the inverse fourth root of this ratio, the resonant response improves it by a factor of 2.8--6.

\onecolumngrid
\begin{center}{\large\bfseries Supplemental Material}\end{center}
\twocolumngrid
\setcounter{equation}{0}
\setcounter{figure}{0}
\renewcommand{\theequation}{S\arabic{equation}}
\renewcommand{\theHequation}{S.\arabic{equation}}
\renewcommand{\thefigure}{S\arabic{figure}}

\section*{S1. From the slab model to a stack of wires}

The homogeneous-slab model of End Matter A captures the resonant enhancement and the competition between radiation and absorption, but it rests on two approximations that our designs satisfy only partially, since successive grids are separated by an appreciable fraction of a wavelength and each grid is a row of discrete wires rather than a uniform sheet. We first examine the role of the grid separation with the impedance-sheet model, and then test the sheet description against a calculation that resolves the individual wires.

The effect of the separation is already visible in an infinite periodic stack of lossless sheets, whose transfer matrix over one period gives~\cite{Brillouin:1946,Collin:1991}
\begin{equation}
\cos Kd=\cos kd+\frac{\pi d}{a\ell}\frac{\sin kd}{kd},
\end{equation}
where $K$ is the Bloch wave number, $k=\omega$ the vacuum wave number, and $\ell=\ln(a/2\pi r)$ the geometric factor of a grid of pitch $a$ and wire radius $r$. In the limit $kd\ll1$ the lowest band edge, where $K=0$, reduces to the plasma frequency of Eq.~\eqref{eq:plasma}, but our grids are about a quarter wavelength apart, $kd\simeq1.7$, and the exact edge then lies 8--23\% lower, for instance at $0.867\,\omega_p$ for $d=a$ and $r/a=0.03$, which is why the designs are computed with the discrete-sheet model rather than with Eq.~\eqref{eq:plasma}. The stack nevertheless keeps its ENZ character, because just above the edge the Bloch phase per period remains small, $Kd\simeq0.27$ for the 1--2\,meV design, so that the field varies slowly from grid to grid even though the vacuum phase across each gap does not.

Then, to test the sheet approximation we resolve the individual wires with the multipole method developed for stacks of cylinder gratings~\cite{Nicorovici:1995,Botten:2000}, in which the field scattered by each wire is expanded in outgoing cylindrical harmonics $H_n^{(1)}(k\rho)e^{in\phi}$, kept up to $|n|\le3$, whose coefficients follow from the impedance condition $E_\parallel=Z_sH_\phi$ on the wire surface, with $Z_s=r_s(1-i\sqrt3)$ combining the normalized surface resistance of the main text with the reactive phase of the anomalous skin effect~\cite{ReuterSondheimer:1948}. Each wire is driven by the fields of all the others in its grid, summed with lattice sums~\cite{Linton:1998}, while different grids are coupled through their plane-wave diffraction orders, including the evanescent ones, since closely spaced grids also interact through their near fields, and the mirror is represented by image wires. The response of the stack then follows by combining the reflection and transmission matrices of the individual grids, as in the transfer-matrix calculation of the sheet model. The calculation is converged, since raising the harmonic cutoff changes the response negligibly even at the most compressed setting, where neighbouring grids are closest and their near-field coupling strongest, and an independent finite-difference frequency-domain calculation for a lossless three-grid stack agrees with it to better than 1\% in the resonance frequency and, after extrapolation in mesh size, in the peak boost, and to about 3\% in the linewidth.

Across the fifteen settings of Fig.~\ref{fig:tuning}(b), resolving the wires lowers the resonance by 0.4--4.7\%, a shift that is compensated by retuning the spacing, after which the coupling reach changes by $-2\%$ to $+4\%$. The forecasts include this correction, interpolated in $\log m_a$.

Finally, the slab response of Eq.~\eqref{eq:betaslab} assumes that the source acts only inside the stack, whereas for DPs, and for axions wherever $\mathbf B_0$ extends in front of the stack, the field also drives the vacuum in front of it. Since this region cannot resonate, it only adds the dish-like emission of the front face, $\delta\beta=\tfrac12(1-r_L)$, where $r_L$ is the reflection amplitude of the stack referred to that face, so that a bare mirror gives $\delta\beta=1$ and any passive stack, with $|r_L|\le1$, gives $|\delta\beta|\le1$, to be compared with a resonant response $|\beta|\simeq25$--75 at the cartridge centers. Including this term in the one-dimensional model lowers $\mathcal I$ by less than 7\%.

\section*{S2. From an infinite lattice to a collected signal}

The response computed so far assumes grids of infinite extent, while a real tile is 30--200 wavelengths across and is read out through optics that accept a single spatial mode. Two effects separate the periodic prediction from the signal a detector actually receives. The finite edges modify the field that the stack radiates, and the receiver collects only the part of that field which overlaps its own mode. We estimate both with a dedicated validation geometry, a 13-grid stack resonant at 1.131\,meV that is distinct from the benchmark cartridges of Table~\ref{tab:design}. In the plane transverse to the wires, each wire is represented by an outgoing cylindrical monopole that satisfies the surface-impedance condition, and the finite mirror by piecewise-constant currents. The wires end on conducting frames, and the variation of the field along them is treated in the paraxial approximation, which is the main approximation of this calculation. Retaining only the monopole is accurate here, since in the periodic limit it changes the peak boost by 0.3\% and the resonance frequency by less than $10^{-4}$ relative to the higher-harmonic calculation of S1.

For a square tile of side 10\,cm the calculated forward power is about 96\% of the periodic prediction, so the edges themselves cost little. The larger reduction comes from collection into a single mode. Even a perfectly uniform square emitter couples only about 79\% of its power into the best-matched Gaussian mode, simply because a flat-topped aperture field is not Gaussian. Projecting the calculated finite-tile field onto this mode gives 76\% of the periodic-model power at 1.131\,meV, and 74--78\% at the additional settings tested near 0.8 and 1.6\,meV. The finite tile therefore behaves almost like an ideal uniform aperture, and the distinction that matters for the forecast is the one between the total emitted power and the power collected in one mode.

An independent vector mode-matching calculation for impedance sheets in a flanged $10\times10$\,cm enclosure leads to the same conclusion. We define $\mathcal I_G$ as the scan integral built from the response collected in the Gaussian mode, and find $\sqrt{\mathcal I_G/\mathcal I}=0.72$--0.78 over the sampled settings between 0.5 and 4\,meV. This ratio can be read as an effective power-collection efficiency, since it equals $\eta$ whenever collection reduces the power by a frequency-independent factor $\eta$. We adopt $\eta_{\rm ap}=0.65$ in the forecasts, somewhat below both calculations, because they support the sampled geometries but do not establish a bound for every cartridge, in particular for the 5\,cm tiles above 4\,meV.

The paraxial treatment along the wires can be tested directly with a three-dimensional finite-difference time-domain (FDTD) calculation, which we provide together with the other codes used in this work~\cite{WiregridGitHub}. The calculation evolves Maxwell's equations on a Yee grid (a standard way to discretize Maxwell's equations for FDTD simulations~\cite{Yee:1966}) for a mirror-backed stack driven by a uniform axion source. The lattice is kept periodic across the wires, while the wires are given their full 10\,cm length and terminated on conducting bars, on conducting walls, or in free space, and each case is compared with infinite wires on the same mesh. For the 1--2\,meV cartridge, where the tile is about 110 wavelengths long, the resonance moves by less than $10^{-4}$ of its frequency, the peak boost and the response width change by at most 3\%, and the collection into a Gaussian mode along the wires agrees with that of a uniform aperture. For the 0.25--0.5\,meV cartridge the tile is only about 30 wavelengths long and the ends matter more. With bars or open ends the collected scan integral is then 10--13\% lower in power than with walls, a reduction that the margin of $\eta_{\rm ap}$ below the calculated collection factors largely absorbs. 

\begin{table}[b]
\caption{Mechanical tolerances at the geometric centers of the five cartridges. The common-gap stability corresponds to a 10\% reduction in power at the intended operating frequency; the rms grid-displacement error corresponds to a 5\% increase in the minimum detectable coupling.}
\label{tab:tolerances}
\begin{ruledtabular}
\begin{tabular}{lcc}
Band [meV] & \shortstack{Gap stability\\{[$\mu$m]}} & \shortstack{rms grid error\\{[$\mu$m]}}\\
\hline
0.25--0.5 & 0.42 & 6.0\\
0.5--1 & 0.33 & 3.8\\
1--2 & 0.26 & 2.4\\
2--4 & 0.21 & 1.4\\
4--8 & 0.13 & 0.7\\
\end{tabular}
\end{ruledtabular}
\end{table}

Dark-matter motion also introduces a further phase variation across the aperture. For a nonrelativistic axion of mass $m_a$ and velocity $\mathbf v$ the driving field carries the wavevector $\mathbf k_a=m_a\mathbf v$, whose component in the tile plane tilts the emitted beam away from the fixed receiver mode. For a uniform zero-velocity aperture field and a receiver mode with real amplitude $u(\mathbf x)$ at transverse position $\mathbf x$, the resulting reduction in collected power is
\begin{equation}
\xi=\frac{\left\langle\left|\int u(\mathbf x)e^{i\mathbf k_a\cdot\mathbf x}\,\dd^2x\right|^2\right\rangle}
{|\int u(\mathbf x)\,\dd^2x|^2},
\label{eq:halo_overlap}
\end{equation}
where the integrals extend over the emitting aperture and the brackets denote an average over the halo velocity distribution. We evaluate these integrals over the square tile with the Gaussian mode truncated at its edges, and $\xi$ enters the forecasts in this form, separately from the finite-tile factor $\eta_{\rm ap}$. Extending the integrals to the full transverse plane for a Gaussian mode with field amplitude proportional to $e^{-|\mathbf x|^2/w^2}$ would give the simpler estimate $\langle e^{-k_\perp^2w^2/2}\rangle$, with $k_\perp$ the wavevector component in the tile plane and $w$ the waist radius. That estimate is somewhat more pessimistic, because truncation at the tile edge narrows the effective aperture and makes it less sensitive to the tilt.

The forecast assumes a Gaussian halo with one-dimensional velocity dispersion $220/\sqrt2$\,km/s and places the 232\,km/s mean wind in the plane of the tile, the orientation that maximizes the phase gradient across the aperture. For 10\,cm tiles this gives $\xi=0.96$ at 1.41\,meV and 0.76 at 4\,meV, and for the 5\,cm tiles $\xi=0.76$ at 8\,meV. With the wind normal to the tile the same values become 0.98, 0.88 and 0.88. The plotted reach therefore corresponds to the least favorable orientation, and averaging the changing projection over a realistic observing schedule would improve the coupling reach by at most 7\% at the highest masses.

Imperfections of the converter itself reduce the collected signal in a similar way, and mechanical precision enters through two distinct requirements, which Table~\ref{tab:tolerances} lists separately. A common change of all gaps, including the one between the mirror and the first grid, only tunes the resonance, so it must merely remain stable while a setting is being measured, and we define its tolerance as the change that lowers the power at the intended operating frequency by 10\%. Errors in the relative positions of the grids instead distort the resonant field, so that different regions of the aperture radiate with different amplitudes and phases and, as for the halo velocity, less power reaches the receiver mode. We model them as independent Gaussian displacements of each grid from its nominal plane, average the complex response over 400 realizations, and define the tolerance as the rms error that raises the coupling required for detection by 5\%. Beyond this point the degradation grows gradually, so that the $5\,\mu$m rms errors already achieved over 100\,mm diameters~\cite{Nagler:2015xta} would raise the required coupling by about 4, 9 and 19\% in the three lowest cartridges and by factors of about 1.5 and 2.4 in the two highest.

\section*{S3. The boost as an overlap with the receiving mode}

End Matter A obtains $\beta$ by solving for the field that the axion drives in the stack and reading off the outgoing wave. The same number follows by asking how the stack \emph{receives} a wave sent toward it from the detector. Then, $\beta$ becomes an overlap integral, which shows directly in the field pattern why a resonant ENZ stack radiates coherently and why other resonances do not.

Let us switch off the axion and let a plane wave of unit amplitude, $e^{-ik(s-L)}$, polarized along the wires, fall on the stack from the detector side. The steady-state field it produces is the receiving mode $E_{\rm rec}(s)$. It obeys the same conditions as the axion-driven field, that is to say it vanishes at the mirror, is continuous at each grid, and its derivative jumps by the sheet current. Outside the stack it is the incident wave plus a reflected one, $e^{-ik(s-L)}+r\,e^{ik(s-L)}$.

\begin{figure}[t]
\includegraphics[width=\columnwidth]{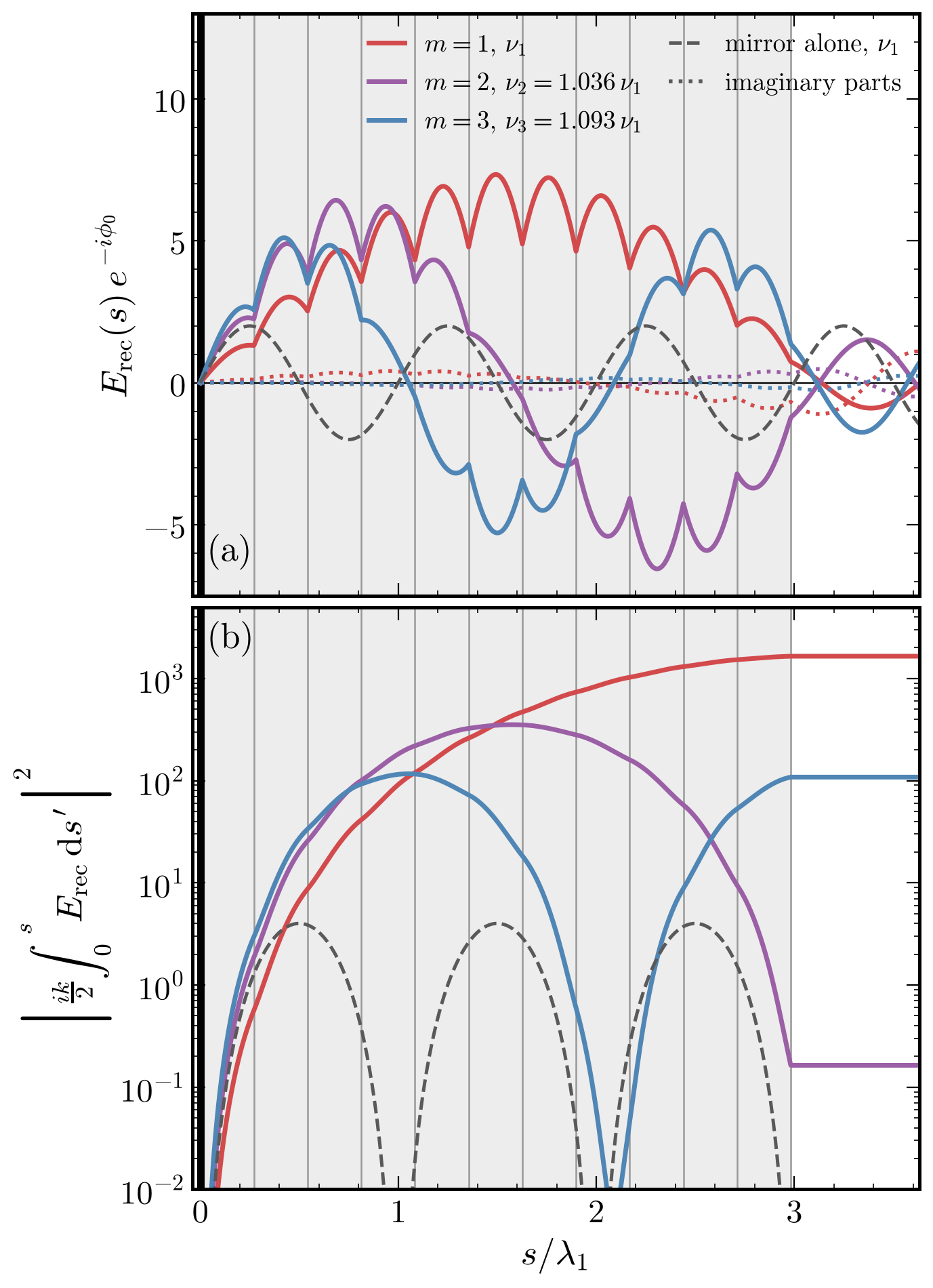}
\caption{Receiving mode of the 1--2\,meV cartridge of Table~\ref{tab:design} ($N=11$, $d=a=238\,\mu$m, $r=0.03a$, $L=2.6$\,mm, impedance-sheet model). The distance from the mirror, $s$, is in units of the vacuum wavelength $\lambda_1=0.88$\,mm at the fundamental resonance $\nu_1=342$\,GHz ($m_a=1.41$\,meV). The thick line marks the mirror, thin lines the grids and the shaded band the stack; the incident wave has unit amplitude. (a)~Real (solid) and imaginary (dotted) parts of $E_{\rm rec}$, after removing a constant phase $\phi_0$, for the first three resonances of the stack, $m=1,2,3$, and real part for the mirror alone. The imaginary part describes the small flow of power into the copper. For $m=1$ it is at most about 12\% of the real part, and $|\int_0^LE_{\rm rec}\,\dd s|/\int_0^L|E_{\rm rec}|\,\dd s=0.99$, so the fundamental mode is a standing wave of nearly uniform phase. (b)~The overlap integral of Eq.~\eqref{eq:overlap} with its upper limit at $s$. Beyond $s=L$ there is no source and the curves give $|\beta|^2$.}
\label{fig:overlap}
\end{figure}

The axion-driven field obeys $E''+k^2E=-k^2E_0$ inside the magnetized stack and $E''+k^2E=0$ outside, where $-E_0$ is the vacuum particular solution of End Matter A. The Green's function of this problem is built from two source-free solutions, the receiving mode $E_{\rm rec}$, which satisfies all the conditions at the mirror and at the grids, and the outgoing wave $e^{ik(s-L)}$, whose Wronskian, evaluated outside the stack, is $2ik$. For $s>L$ this gives $E(s)=\beta E_0e^{ik(s-L)}$ with
\begin{equation}
\beta=\frac{ik}{2}\int_0^L E_{\rm rec}(s)\,\dd s ,
\label{eq:overlap}
\end{equation}
where the integral runs over the region in which the axion source acts. The meaning of this result is simple. A thin slice $\dd s$ of the uniform source acts as a small antenna, which in empty space would radiate a wave of amplitude $(k\,\dd s/2)E_0$ toward the detector, and in the presence of the grids and the mirror reciprocity implies that this wave reaches the detector with the amplitude and phase of $E_{\rm rec}$ at the position of the slice. The boost is therefore the sum of the contributions of all slices, each weighted by the receiving mode, and it is large only if this mode is both strong inside the stack and of the same phase throughout, so that the contributions add rather than cancel. For this reason, the loaded quality factor and coupling measured by reflectometry do not by themselves determine the axion sensitivity of a multilayer, which also depends on the sign structure of the receiving mode~\cite{Garutti:2026asa}. As a check, for a mirror alone Eq.~\eqref{eq:overlap} gives $\beta=(1-\cos kL)e^{ikL}$, the $\varepsilon=1$ limit of Eq.~\eqref{eq:betaslab}, and for the cartridges of Table~\ref{tab:design} it reproduces the direct solution of End Matter A to numerical precision. Equation~\eqref{eq:overlap} is the one-dimensional form of the overlap formalism for dielectric haloscopes~\cite{Ioannisian:2017srr,Millar:2016cjp} and of the reciprocity approach of Refs.~\cite{Egge:2022gfp,Egge:2023cos}.

In front of a bare mirror, the receiving mode is the familiar standing wave, which changes sign every half wavelength (dashed curves in Fig.~\ref{fig:overlap}), so that neighboring half-wavelength layers contribute with opposite signs and cancel in pairs. However thick the converting region, at most one uncompensated layer remains, and $|\beta|^2$ never exceeds 4.

The stack instead behaves like a string fixed at both ends. The field vanishes at the mirror and, at a resonance, nearly vanishes at the front face as well, because the internal amplitude produced by a unit incident wave, about $2/n_{\rm eff}$ with $n_{\rm eff}$ the effective refractive index of the lattice, is largest when the face sits at a node of the internal standing wave. The resonances are therefore standing waves that fit an integer number $m$ of half internal wavelengths across the stack, $KL\simeq m\pi$, with $K$ the wave number inside the lattice, so that mode $m$ has $m-1$ nodes, as shown for the first three in Fig.~\ref{fig:overlap}(a). Only the fundamental mode keeps the same sign throughout the stack, so that every slice of the source adds in phase and the overlap in panel~(b) grows steadily up to the peak boost of Table~\ref{tab:design}. In the antisymmetric mode $m=2$, by contrast, the two lobes of opposite sign cancel almost exactly, so that the axion barely excites it even though it appears clearly as an absorption dip in reflection, while for $m=3$ two of the three lobes cancel each other, leaving a power boost about fifteen times smaller than that of the fundamental. This cancellation between lobes is the origin of the factor $1-\cos\phi$ in Eq.~\eqref{eq:betaslab}, vanishing for even $m$.

The coherent fundamental mode thus requires the stack to contain only half an internal wavelength. In vacuum, this would limit the converting region to $\lambda/2$, the thickness of the first lobe of the mirror's standing wave, whereas the stack of Fig.~\ref{fig:overlap} is three vacuum wavelengths thick, which requires a wavelength inside about six times longer than in vacuum and hence a permittivity of a few percent. This is the ENZ condition, seen from the receiver.

\end{document}